\documentclass[journal,onecolumn]{IEEEtran}

\usepackage{multirow}
\usepackage{amsfonts}
\usepackage[mathcal]{eucal}
\usepackage{mathtools}
\usepackage{color}
\usepackage{algpseudocode}
\usepackage{algorithm}
\usepackage{program}
\usepackage{amsmath}
\usepackage{graphics}
\usepackage{epstopdf}
\usepackage{balance}
\usepackage{subcaption}
\usepackage{cite}
\usepackage{array}
\newcolumntype{P}[1]{>{\centering\arraybackslash}p{#1}}
\newcolumntype{M}[1]{>{\centering\arraybackslash}m{#1}}

\graphicspath{{/}}
\newcommand{\paren}[1]{\left(#1\right)}
\newcommand{\sqparen}[1]{\left[#1\right]}
\newcommand{\brparen}[1]{\left\{#1\right\}}
\newcommand{\field}[1]{\ensuremath{\mathbb{#1}}}
\newcommand{\N}{\ensuremath{\field{N}}} 
\renewcommand{\R}{\ensuremath{\field{R}}} 
\newcommand{\I}[1]{\ensuremath{\mathsf{1}_{\left\{#1\right\}}}} 
\newcommand{\PRP}[1]{\ensuremath{\mathsf{Pr}\left(#1\right)}} 
\newcommand{\vecbold}[1]{\ensuremath{\boldsymbol{#1}}}

\newcommand{\floor}[1]{\lfloor #1 \rfloor}

\newtheorem{theorem}{Theorem}

\newtheorem{definition}{Definition}

\DeclareMathOperator*{\argmax}{\arg\!\max}

\begin{document}
%
\title{Optimal Network-Assisted Multiuser DASH Video Streaming}
%
%
\author{Emre~Ozfatura,
        Ozgur~Ercetin,
        Hazer~Inaltekin
\thanks{ E. Ozfatura is with the Department of Electrical Electronic Engineering, Imperial College London, London SW7 2AZ, UK.} 
\thanks{ O. Ercetin is with the Faculty of Engineering and Natural Sciences, Sabanci University, 34956 Istanbul, Turkey.}
\thanks{ H. Inaltekin is with the Department of Electrical Engineering, Princeton University, Princeton, NJ 08544, USA.}
\thanks{This work is supported in part by a grant from Argela Technologies, Turkey.}}
\maketitle

\begin{abstract}
Streaming video is becoming the predominant type of traffic over the Internet with reports forecasting the video content to account for 82\% of all traffic by 2021. With significant investment on Internet backbone, the main bottleneck remains at the edge servers (e.g., WiFi access points, small cells, etc.).  In this work, we obtain and prove the optimality of a multiuser resource allocation mechanism operating at the edge server that minimizes the probability of stalling of video streams due to buffer under-flows.  Our derived policy utilizes Media Presentation Description (MPD) files of clients that are sent in compliant to Dynamic Adaptive Streaming over HTTP (DASH) protocol to be cognizant of the deadlines of each of the media file to be displayed by the clients.  Our policy allocates the available channel resources to the users, in a time division manner, in the order of their deadlines. After establishing the optimality of this policy to minimize the stalling probability for a network with links associated with fixed loss rates, the utility  of the algorithm is verified under realistic network conditions with detailed NS-3 simulations.
\end{abstract}


\section{Introduction}
\label{sec:introduction}
\IEEEPARstart{T}{here} is an increasing demand for multimedia streaming applications thanks to the ubiquity of internet access, the availability of the online content and the growing number of wireless hand-held devices. The predictions of Cisco Visual Networking Index \cite{cisco} indicate that IP video traffic will constitute 82 percent of all consumer internet traffic by 2021. For instance, in 2016, YouTube and Netflix account for up to 53 percent of fixed access Internet traffic in North America \cite{sandvine}. Moreover, 21 percent of the mobile internet traffic in North America is solely based on YouTube \cite{sandvine}.

\begin{figure*}
	\centering
		\includegraphics[scale=0.4]{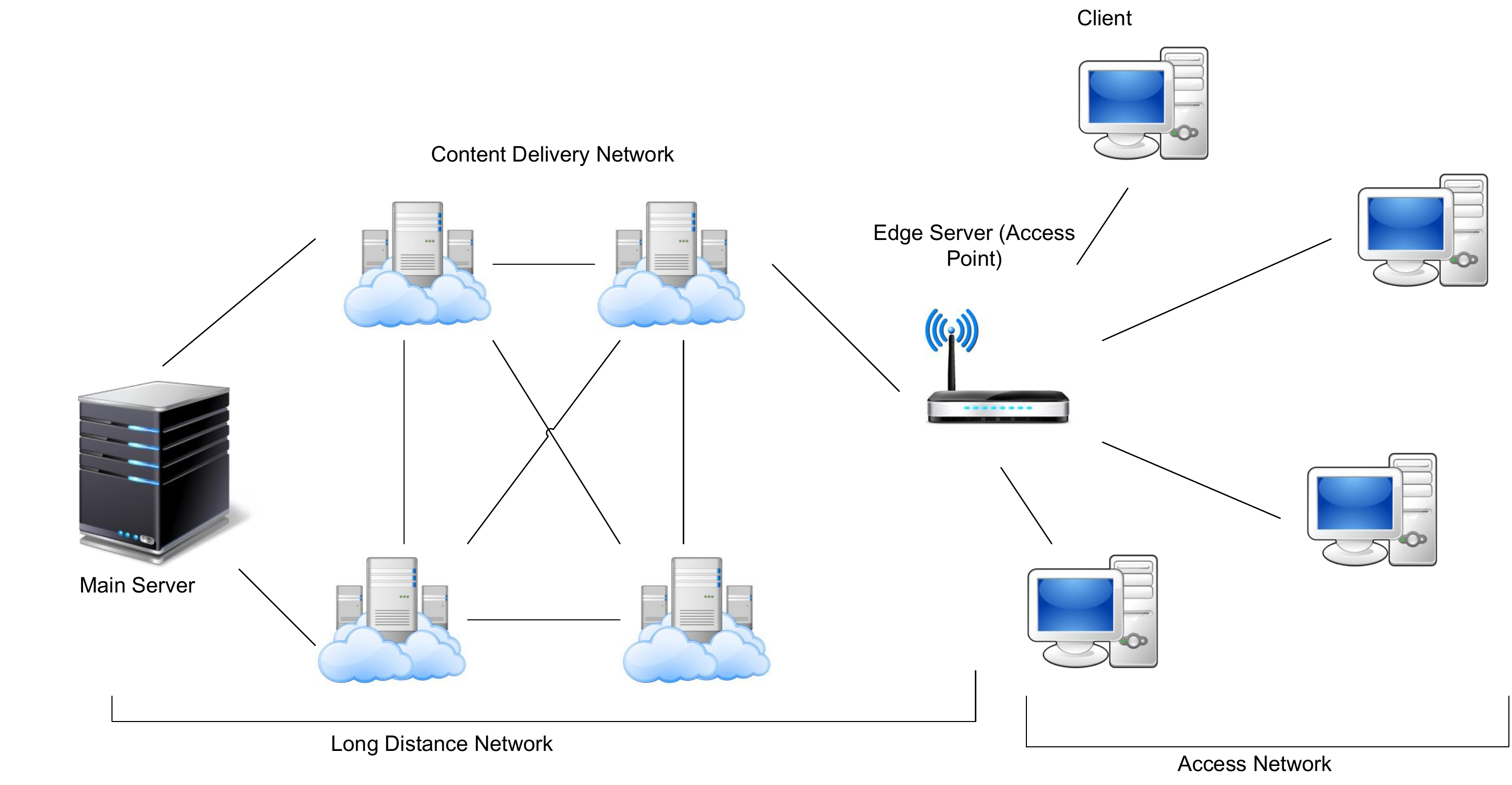}
     \caption{Multiuser video streaming system}
	\label{fig1}
\end{figure*} 

In this work, we derive a Dynamic Adaptive Streaming over HTTP (DASH)-compatible multiuser resource control policy, which we call  blind deadline-based resource allocation (BDRA) scheme, operating at an edge server.  The aim of the BDRA scheme is to perform slot-based resource allocation to users in order to minimize the probability of a \emph{stalling} event at a client.  When the amount of data at the buffer of a client is insufficient to continue to display the video stream, a stalling event occurs, and the client begins a re-buffering period during which it fills its buffer without displaying the video stream. The BDRA scheme utilizes Media Presentation Description (MPD) files of clients and HTTP-GET requests, which are sent in compliant to DASH protocol, in order to define and update the deadline of each media file displayed by a client. Then, it allocates  slots to the users in the order of their deadlines.  We formally prove that this algorithm minimizes the stalling probability for a network with links associated with fixed loss rates.

In conventional applications of  DASH framework, the client is the only agent that manages the video streaming process in order to maximize the subjective video quality \cite{csca1,csca2,csca3,csca4,csca5,csca6}. In particular, the main promise of DASH is that the clients dynamically select among different representations of the same media stream differing with respect to video encoding rates based on the estimated network throughput. However, while each client  has access to only its own MDP file, the edge server has access to MDP files of all clients it is serving. Hence, the edge server has a better view of the overall operation of the network, and it is in a position to \emph{proactively} take resource allocation decisions to prevent stalling events, whereas individual clients can only react after a stalling event occurs.
%
%
%
%

The DASH protocol has several benefits over push-based media streaming protocols such as Real-time Transport Protocol (RTP) \cite{dash1,dash2,dash3,dash4}. First, the infrastructure of the Internet has evolved to efficiently support HTTP, and HTTP offers ubiquitous connectivity. 
Second, DASH is a pull-based protocol, so it traverses the firewalls. Third, the underlying TCP/IP protocol is widely deployed and provide reliable data transmission. Fourth, a client does not have to maintain a session state on the server to stream over the HTTP reducing overhead at the server. 

The importance of the problem is well established as indicated by past and ongoing studies, e.g., \cite{main1} and \cite{main2}. In \cite{main1}, the authors examined the joint optimization of network resource allocation and video quality adaptation.  The authors propose a resource allocation algorithm that aims to prevent the stalling event by employing a parameter that reflects the risk of stalling according to the duration of the video in the  client’s buffer. A larger rate is assigned to a user that has a high stalling risk.     
In \cite{main2}, the authors introduce the notion of {\em playout lead}, which is defined as the duration of the additional time a client can play the video by using its currently buffered data. The authors propose an algorithm that aims to prevent  the stalling occurrences  by maximizing the playout lead for all clients. To this end, the resource (time slots) is allocated so that the minimum of the playout lead among all users is maximized.  Besides \cite{main1} and \cite{main2}, a buffer-aware approach is considered in \cite{bufferaware} where the video streaming traffic is shaped by SDN controllers according to clients' buffer status and the buffer occupancy trends. 

Our work improves the current state-of-the-art in two ways. First, we prove that the derived policy of serving the clients in the order of deadlines is optimal in the sense that it minimizes the stalling event probability of the network when the link loss rates are fixed.\footnote{ When the instantaneous channel state information is available, this information can be further utilized to modify the scheduling algorithm in order to improve the network performance as in \cite{channelbased}.}  Second, our policy relies only on the acknowledgment (ACK) feedback from the clients taken in the form of HTTP-GET requests for the subsequent byte ranges of the media file, and thus, significantly reducing the implementation complexity.

Note that in HTTP adaptive streaming, the quality of experience (QoE) depends on the selection of different system parameters such as initial setup delay, re-buffer duration, average video quality,  video quality fluctuations and the number of stalling events \cite{vq1,vq2,vq3,vq4,vq5,vq6}. In practical DASH implementations, on the other hand, the client is only responsible for the video quality selection process. In this paper, considering the initial setup delay and re-buffer duration as predefined system parameters, we focus on minimizing the stalling probability with our server side algorithm in order to improve the QoE of users. To the best of our knowledge, this paper is the first study that proposes a systematic approach based on Markov decision processes (MADP) in order to investigate the performance of DASH based multiuser video streaming systems with network assistance. The proposed MADP framework enables us to take into account the effect of resource allocation decisions in the current time-slot on the stalling likelihood in future time-slots over a finite time horizon. Our main contributions in the paper are summarized as follows.
\begin{itemize}
\item Using dynamic programming, we show that the {\em optimum} algorithm minimizing the system-wide stalling probability in DASH based multiuser video streaming systems when only statistical knowledge of channels is available at the server-side is a {\em blind} deadline based algorithm, which we call the BDRA algorithm.

\item Having a simple structure with polynomial-time computational complexity, the BDRA algorithm is easy to implement as a server-side add-on solution for the existing DASH architecture in order to reduce the frequency of stalling events. We further provide a particular implementation of the BDRA algorithm that prioritizes the users with small GoP sizes in order to achieve fairness among the streaming users with varying bit-rates for video files. 
\item Thanks to its operation oblivious to the quality adaptation mechanism at the client side, the BDRA algorithm can operate together with any choice of quality adaptation scheme such as buffer-based adaptation (BBA) and rate-based adaptation (RBA), which further increases the utility of the derived BDRA algorithm. 
\item We perform NS3 simulations in order to illustrate the {\em optimality gap} between the BDRA scheme and four other blind resource allocation schemes.   
\end{itemize}


%
%

The remainder of the paper is organized as follows. In Section \ref{sec:background}, we provide a detailed background on the operation of the DASH protocol. 
Section \ref{Section: System Model} provides the analytical model for our system as well as the optimum slot-based resource allocation problem to be solved. The BDRA algorithm is formally introduced and its optimality is formally established in Section \ref{sec:optimum}. 
Implementation and design issues regarding the BDRA algorithm are explained in  Section \ref{sec:Implementation}. Performance of the BDRA algorithm, in comparison to commonly used {\em rate-fair resource allocation} schemes, is numerically investigated in Section \ref{sec:numerical}.  Section \ref{sec:related} provides a detailed discussion on the previous work that is most relevant to our findings in this paper, by first describing the current state-of-the-art and then explaining the differences between our solution and these previous solutions in detail. Finally, we conclude the paper with a summary of findings and future research directions in Section \ref{sec:conclusion}.

\section{DASH Video Streaming}
\label{sec:background}

As illustrated in Fig. \ref{fig1}, the studied video streaming system consists of two main sections: A Long Distance Network (LDN) and an Access Network (AN). The LDN may involve both a main server and a content delivery network (CDN), and it has the responsibility of delivering the requested video files to the edge servers in the AN. In general, the bottleneck of the end-to-end connection is at the edge servers, so we focus on the resource allocation strategies operating at the edge servers to alleviate this bottleneck.

Due to their significant advantages over push-based media streaming protocols such as RTP, HTTP-based streaming protocols have been widely adopted by most of the on-demand video service providers. In particular, DASH protocol is developed to provide a common set of functionality among different HTTP-based streaming protocols \cite{dash1,dash2,dash3,dash4}.  

In DASH, a video file is encoded with multiple different bit-rates into different \emph{representations}, where each representation corresponds to a different level of quality of the same video stream.   Each representation is broken into \emph{segments} of duration 2-10 seconds \cite{seg}. Segments may be further subdivided into \emph{sub-segments}, each of which contains a whole number of complete access units. The video content providers employing DASH  often use video files encoded according to Advanced Video Coding (AVC) (e.g., H264.AVC) standard. In this video encoding format, the smallest meaningful bit-chunk is called {\em Group of Pictures} (GoP) since the frames of the same GoP are encoded and decoded together \cite{gop,dash4}. Thus, an AVC encoded video file is considered as a combination of mutually exclusive fragments that contain different frames of the same video file. Each GoP contains a fixed number of frames and has a fixed video display duration.  We note that although each GoP has a fixed video display duration, their sizes might be different due to video content.  To display a frame, all information related to the corresponding GoP needs to be available at the client buffer. \\
\begin{figure*}
	\centering
		\includegraphics[scale=0.5]{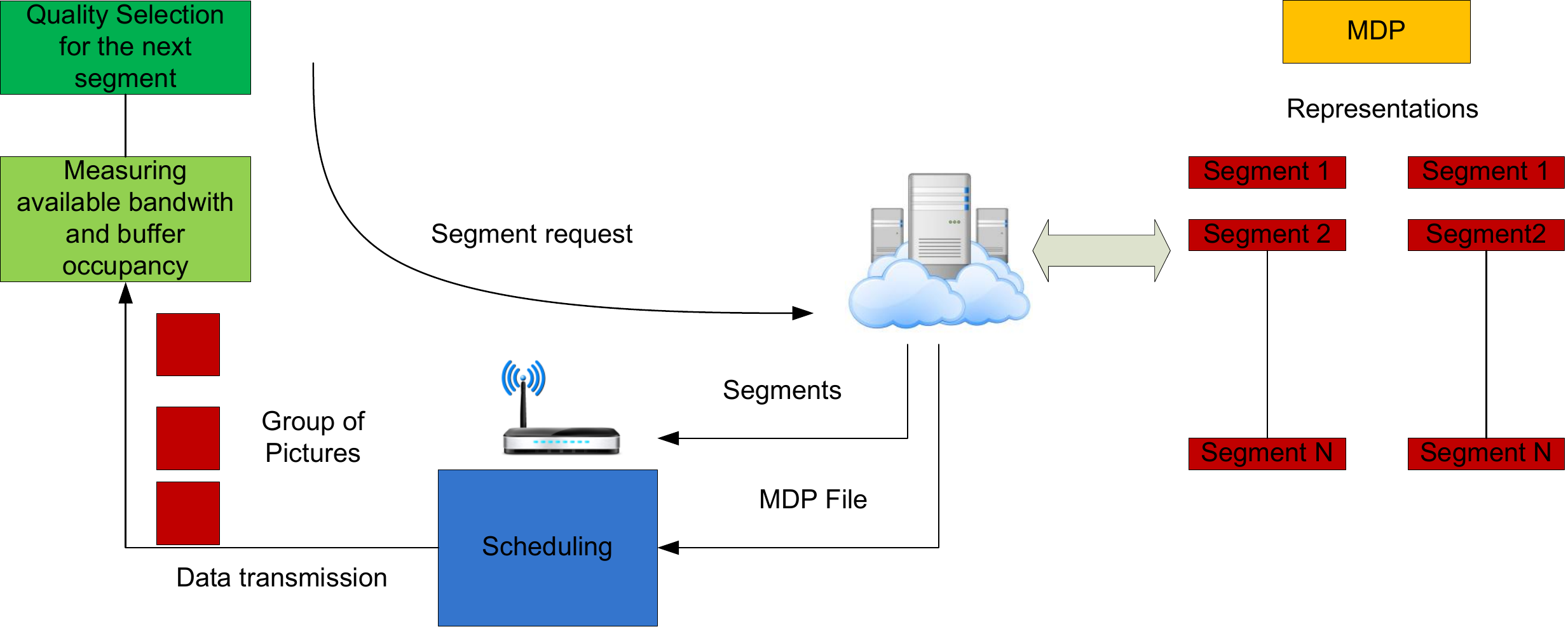}
		\caption{ End-to-end video streaming system.}
	\label{fig2}
\end{figure*}
\indent 
The DASH client behavior can be summarized as follows.  The client first accesses the Media Presentation Description (MPD) file. The MPD file contains metadata required by a DASH client to construct appropriate HTTP-URLs to access segments and to provide the streaming service to the user. In particular, an MPD file provides information for the earliest presentation time and presentation duration for each segment in the representation.\footnote{This information will be used by the derived optimum algorithm to perform resource allocation among multiple DASH clients.} The client selects an appropriate video representation, typically based on an estimate of the available bandwidth to the server but also on the rendering capabilities of the client.  Then, the client creates a list of accessible segments for each representation. The client accesses the content by requesting entire segments or byte ranges of segments via HTTP-GET command. Once the presentation has started, the client continues consuming the media content by continuously requesting segments or parts of segments.  
The client may switch representations taking into account  updated information from its environment, e.g., change of observed throughput. In this paper, we focus on the  resource allocation at the edge server.  Hence, DASH clients can use any adaptive video quality selection algorithm to select an appropriate representation based on the observed throughput and client capabilities.


%
%
%

\section{Analytical Model, Definitions and the Optimum Scheduling Problem } \label{Section: System Model}
In this section, we will introduce the details of our analytical model (following the standard terminology of the {\em MADP}  literature \cite{dynamicprog}), the definitions that go with this model and the {\em optimum scheduling problem} that we solve to minimize  the number of stalling events in DASH based multiuser video streaming systems.

\subsection{Receiver and Playout Curves}

 The data arrival process of client $i$ is denoted by $R_{i}(t)$, which we call the \emph{receiver curve} of client $i$. The receiver curve $R_{i}(t)$  indicates the total amount of error free data in unit of packets that is delivered to client $i$ up to time $t$. For each client $i$, $R_{i}(t)$ is a non-decreasing function of $t$.  The video of client $i$ is displayed according to $p_{i}(t)$, which is called the \emph{playout curve}. The playout curve describes the minimum amount of data in units of packets that needs to be decoded up to time $t$ to perform uninterrupted video display. 
 \begin{figure}
  \centering
	\includegraphics[scale=0.4]{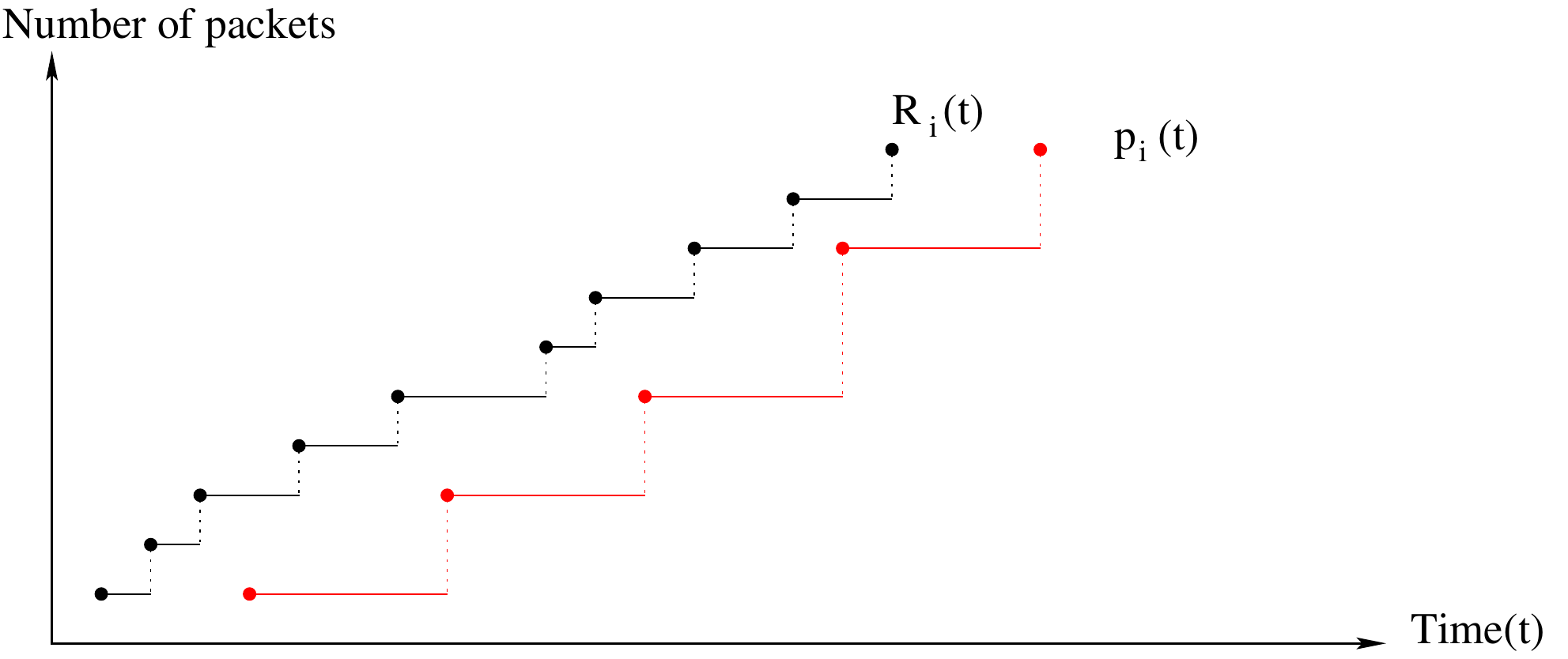}
	\caption{Receiver and Playout curves}
	\label{systemparameter}
\end{figure}
The GoP based structure of the video files implies that all playout curves are right continuous functions as illustrated by Fig. \ref{systemparameter}. A time instant $t>0$ at which there is a jump in the playout curve, i.e., $p(t^-) \neq p(t)$ for any $t^- < t$, is called an {\em increment point}. 

We consider a time-slotted video streaming system with fixed slot length equal to $\Delta$ so that the edge server can serve only one user in each slot duration. Hence, the receiver curve $R_{i}(t)$ increases by one unit  at the end of a time-slot if and only if user $i$ is scheduled at the beginning of the corresponding time-slot and the transmitted packet is received successfully. As a result of this operation, all receiver curves are right continuous functions as well, an example of which is illustrated in Fig. \ref{systemparameter}. We further assume that GoP duration is also an integer multiple of the slot duration $\Delta$. Since both playout curves and receiver curves remain constant during a slot duration, we can discretize these functions and use time index $k=\floor{t/\Delta}$, where  $\floor{\cdot}$ is the {\em floor} function that produces the largest integer smaller than or equal to its argument. Throughout the paper, we will normalize $\Delta$ to {\em one} time unit to simplify notation. To ensure continuous displaying of a video at client $i$, there should be sufficient number of packets in the client buffer so that the following inequality holds
\begin{equation}\label{neccond}
R_{i}(t) \geq p_{i}(t)
\end{equation}
for any time instant $t$.
\subsection{Analytical Model and Definitions}
Our primary aim is to discover the structure of the optimum scheduling policy (at the edge server side) that will minimize the stalling event probability for multiuser video streaming over stochastically varying wireless channels.  To this end, we focus on minimizing the stalling probability per segment, where each segment spans $T \in \N$ consecutive slots of time.\footnote{The main reason for us to consider {\em only} the segment stalling probability in this paper is the technological constraint introduced by the DASH protocol.  In particular, the DASH protocol determines the representation level of the next segment only after the current segment requests are provisioned, and we cannot state our optimum scheduling problem without knowing the representation levels of the forthcoming segments.} Hence, without loss of generality, we model our optimum slot-based resource allocation problem as a {\em finite} horizon stochastic dynamic programming problem over time interval $[0, T]$ below.  

The classical packet {\em erasure} channel is used to model wireless channels between the end users and the edge server, as such a packet sent for user $i$ is either successfully received with probability $\beta_i$ or lost with probability $1-\beta_i$ in each time slot.  
We assume that channel statistics $\vecbold{\beta}=(\beta_{1},\ldots,\beta_{N})$ are known  at time $t=0$ and remain the same over the time interval $\sqparen{0, T}$. Similarly, we also assume that playout curves (or, alternatively called representation levels) $\vecbold{p}(t)=\paren{p_{1}(t),\ldots, p_{N}(t)}$ are known by the edge server at time $t=0$, which is a standard assumption of the DASH protocol. Here, $\vecbold{p}(t)$ is a vector valued function that describes the amount of data (measured in terms of number of packets) required by each user up to time $t$ to display its video without any interruptions.       


The edge server can serve {\em only one} user in each time slot.  Hence, a scheduling decision must be made at the beginning of each time slot to select an {\em appropriate} user (i.e., usually the one that optimizes the system performance) for data transmission based on the current system state that summarizes the data reception history. In this paper, we represent the system states by the $N$ dimensional vector $\vecbold{s}=\paren{s_{1},\ldots,s_{N}}$, where $s_{i}$ is equal to the number of packets received by user $i \in \brparen{1, \ldots, N}$. We will often use states with time index $\floor{t}$ (or, by using the discrete time index $k \in \brparen{0, \ldots, T-1}$), i.e., $\vecbold{s}_{\floor{t}}=\paren{s_{1,\floor{t}},\ldots,s_{N,\floor{t}}}$, to denote the number of packets received by the users at the beginning of time slot $\floor{t}$. The set $\mathcal{S} = \brparen{0,1,2,\ldots,T}^{N}$ defines the set of all state vectors.\footnote{We note that $\mathcal{S}$ is larger than the set of all admissible states. If needed to be more precise, we can write $\mathcal{S}^{'}=\brparen{\vecbold{s}\in\mathcal{S}: \sum_{i=1}^N s_{i}\leq T}$.} 


In this setting, we define the {\em action set} $\mathcal{A}$ to be $\mathcal{A}= \left\{1,2,\ldots,N\right\}$, and each action $a$ belonging to $\mathcal{A}$ denotes the index of the user scheduled for video streaming in the current time-slot.  We note that $\mathcal{A}$ is a state-independent action set that remains the same for all $\vecbold{s}\in\mathcal{S}$.  Consider now a specific time-slot $k$.  An important quantity of interest that describes how the video streaming system in question evolves in time is the {\em transition probability function} $P_{k}\paren{\vecbold{z} \vert \vecbold{s}, a}$ that represents the transition probability of the video streaming system to  another system state $\vecbold{z}$ at the beginning of the next time-slot given that the  system state in the current time-slot $k$ is $\vecbold{s}$, i.e., $\vecbold{s}_k = \vecbold{s}$, and the action taken in this time-slot is $a$.  Using the wireless channel model between the edger server and the users, $P_{k}\paren{\vecbold{z} \vert \vecbold{s}, a}$ can be more formally written as   
\begin{eqnarray*}
P_{k}\paren{\vecbold{z} \vert \vecbold{s}, a} = \left\{\begin{array}{cc} 1-\beta_{a} & \mbox{ if } \vecbold{z} = \vecbold{s} \\ \beta_{a} & \mbox{ if } \vecbold{z} = \paren{s_{1},\ldots,s_{a}+1,\ldots, s_{N}}\\ 0 & \mbox{ otherwise} \end{array}\right..
\end{eqnarray*}


In addition to the analytical framework introduced above, two other major components of our model that operate on this framework are {\em decision rules} and the {\em scheduling policy}, which are what we define next. Considering the fact that packet success or failure events are independent from time-slot to time-slot in our wireless channel model\footnote{This assumption implies that knowing the transmission history and associated success or failure events do not give us any information about the channel conditions in the current time-slot.}, knowledge of the current system state is sufficient to predict current channel conditions and to construct remaining playout curves, i.e., remaining demand for data for uninterrupted video streaming. Hence, without loss of generality, we focus on Markovian and deterministic decision rules defined as functions that map the set of states $\mathcal{S}$ to the set of actions $\mathcal{A}$. More specifically, the decision rule $d_k$ for time-slot $k$ takes the system state $\vecbold{s}_k$ in the beginning of this time-slot as an input, and produces an action $a$ belonging to $\mathcal{A}$, i.e., $d_k\paren{\vecbold{s}_k} = a \in \mathcal{A}$, that represents the user index scheduled for video streaming in this time-slot. 

Utilizing the definition of decision rules, we next state the definition of scheduling policy and {\em tail} scheduling policy below, which will conclude the description of our analytical model.    
  
\begin{definition}
A scheduling policy $\vecbold{\pi} = \paren{d_{0}, \ldots, d_{T-1}}$ is a sequence of decision rules as such the $k$th element of $\vecbold{\pi}$ determines the index of the user scheduled for the $k$th time-slot based on the observed system state at the beginning of this time-slot for $k \in \brparen{0, \ldots, T-1}$.  Similarly, a tail scheduling policy $\vecbold{\pi}_k = \paren{d_k, \ldots, d_{T-1}}$ is a sequence of decision rules that determines the indices of the users scheduled for the time-slots from $k$ to $T-1$.       
\end{definition}


\subsection{The Optimum Scheduling Problem}
Having introduced our analytical model above, we are now ready to state the optimum scheduling problem.  To this end, we first need to define {\em total expected reward} that is obtained when the user scheduling policy $\vecbold{\pi} = \paren{d_{0}, \ldots, d_{T-1}}$ is employed to determine scheduling decisions for each time slot.  
\begin{definition}
The total expected reward $u^{\boldsymbol{\pi}}_{k}: \mathcal{S}\mapsto \R$ collected from time-slot $k$ to $T-1$ under the scheduling policy $\vecbold{\pi} = \paren{d_0, \ldots, d_{T-1}}$ is a function that maps the initial system state $\vecbold{s}_k$ at the beginning of the time-slot $k$ to a real number.    
\end{definition}

We note that $u^{\boldsymbol{\pi}}_{k}$ can be easily expressed recursively as 
\begin{equation}\label{receq1}
u^{\boldsymbol{\pi}}_{k}\paren{\vecbold{s}_{k}} = r_{k}\paren{\vecbold{s}_{k}, a} + \sum_{ \vecbold{s} \in \mathcal{S}} P_{k}\paren{\vecbold{s} \vert \vecbold{s}_{k},a} u^{\vecbold{\pi}}_{k+1}\paren{\vecbold{s}}
\end{equation}
for any $\vecbold{s}_k \in \mathcal{S}$, where $r_{k}\paren{\vecbold{s}_{k}, a}$ denotes the reward obtained by the scheduling decision $a = d_{k}\paren{\vecbold{s}_{k}}$ at time-slot $k$ if the current system state is $\vecbold{s}_{k}$, and the summation term in \eqref{receq1} represents the total expected reward obtained from time-slot $k+1$ onwards.  It should be noted that $u^{\boldsymbol{\pi}}_{k}\paren{\vecbold{s}_{k}}$ in \eqref{receq1} depends on $\vecbold{\pi}$ only through its tail policy $\vecbold{\pi}_k = \paren{d_k, \ldots, d_{T-1}}$.  For the sake of completeness, we set $u^{\vecbold{\pi}}_{T}\paren{\vecbold{s}_{T}} = r_{T}\paren{\vecbold{s}_{T}}$, where it is understood that $\vecbold{s}_{T}$ is the system state reached at the end of the video segment of interest, $r_{T}\paren{\vecbold{s}_{T}}$ is the reward collected due to the occurrence of $\vecbold{s}_{T}$, and no action is allowed at this termination time, which is a standard assumption of the finite horizon stochastic control problems \cite{dynamicprog}. The notion of {\em optimality} for a scheduling policy is introduced in the following definition.       
\begin{definition} \label{Def: Scheduling Optimality}
Let $\Pi$ be the set of all scheduling policies. Then, we say that a scheduling policy $\vecbold{\pi}^\star$ is {\em optimum} if it solves the optimization problem below
\begin{eqnarray}
\max_{\vecbold{\pi} \in \Pi} u^{\boldsymbol{\pi}}_{k}\paren{\vecbold{s}} \label{Eqn: Scheduling Optimality}
\end{eqnarray}  
for all time-slots $k \in \brparen{0, \ldots, T-1}$ and initial state vectors $\vecbold{s} \in \mathcal{S}$.\footnote{The maximum value in \eqref{Eqn: Scheduling Optimality} is always achieved since $\Pi$ is a finite set, and hence there is no ambiguity in this definition.}  
\end{definition} 
 
We note that the condition of optimality introduced in Definition \ref{Def: Scheduling Optimality} is a strong one since we do not only want a given scheduling policy is optimum itself considering time-slots from $0$ to $T-1$ but also want all of its tail policies to be optimum and achieve the {\em best} possible total expected reward starting from any time-slot and initial system state.  To put it in another way, we want an optimal scheduling policy $\vecbold{\pi}^\star$ to satisfy the following equality  
\begin{equation}
u^{\vecbold{\pi}^{\star}}_{k}\paren{\vecbold{s}}=u^{\star}_{k}\paren{\vecbold{s}}
\end{equation}
for all $k \in \brparen{0, \ldots, T-1}$ and $\vecbold{s} \in \mathcal{S}$, where $u^{\star}_{k}\paren{\vecbold{s}} = \max_{\vecbold{\pi} \in \Pi} u^{\vecbold{\pi}}_{k}\paren{\vecbold{s}}$.  

We will derive the structure of $\vecbold{\pi}^\star$ by considering a specific but practically relevant total expected reward function, which is the system-wise segment {\em non}-stalling probability, i.e., none of the users experiences stalling throughout a particular segment duration.  Indeed, our problem formulation lends itself to readily calculate the segment non-stalling probability if we set the per-slot reward functions $r_k\paren{\vecbold{s}, a}$ to zero for all $k \in \brparen{0, \ldots, T-1}$, set $r_{T}\paren{\vecbold{s}_{T}}$ to zero (one) if a stalling event does (not) occur at the end of time-slot $T-1$ (i.e., the termination time).\footnote{The condition to check if a stalling event occurs or not at the end of time-slot $T-1$ is equivalent to checking the inequality $s_{i, T} \geq p_i\paren{T}$ for all $i \in \brparen{1, \ldots, N}$. If this inequality is not satisfied for a user, we say that a stalling event occurs at the termination time $T$.}  Accordingly, the total expected reward in \eqref{receq1} for the segment stalling probability can be written as
\begin{eqnarray} \label{receq2} 
\lefteqn{u^{\vecbold{\pi}}_{\floor{t}}\paren{\vecbold{s}_{\floor{t}}}}  \hspace{8cm} \nonumber \\
\lefteqn{ = \left\{\begin{array}{cc}  \sum_{\vecbold{s} \in\mathcal{S}} P_{\floor{t}}\paren{\vecbold{s} \vert\vecbold{s}_{\floor{t}}, a}u^{\vecbold{\pi}}_{\floor{t}+1}\paren{\vecbold{s}} & \mbox{ if } \vecbold{s}_{\floor{t}} \succeq \vecbold{p}(t) \\ 0 & \mbox{otherwise} \end{array}\right.,} \hspace{8cm}
\end{eqnarray}
where ``$\succeq$" represents element-wise vector inequality and $a$ is the action taken in time-slot $\floor{t}$ by the scheduling policy $\vecbold{\pi} = \paren{d_0, \ldots, d_{T-1}}$, i.e., $a =d_{\floor{t}}\paren{\vecbold{s}_{\floor{t}}}$. It should be noted that a given scheduling policy $\vecbold{\pi}$ induces a probability distribution over the set of system states $\mathcal{S}$, which in turn determines a probability distribution for {\em random} receiver curves $R_i\paren{t}$ for $i \in \brparen{1, \ldots, N}$ and $t \in [0, T]$. Hence, $u^{\vecbold{\pi}}_{\floor{t}}\paren{\vecbold{s}_{\floor{t}}}$ can also be written as the probability that all random receiver curves to be above all playout curves over the time interval $[t, T]$ starting from the initial system state $\vecbold{s}_{\floor{t}}$.  That is, $u^{\vecbold{\pi}}_{\floor{t}}\paren{\vecbold{s}_{\floor{t}}}$ is equal to
\begin{eqnarray}
\lefteqn{F_{\floor{t}}\paren{\vecbold{p},\vecbold{\pi}_{\floor{t}},\vecbold{s}_{\floor{t}}}} \nonumber \hspace{8cm}\\ 
= \lefteqn{\PRP{ \bigcap_{i=1}^N \brparen{R_i(\tau) \geq p_i(\tau), \forall \tau \in\sqparen{t, T}} \Big\vert \ \vecbold{s}_{\floor{t}}}.}  \label{Eqn: Stalling Probability} \hspace{7cm}
\end{eqnarray}       
Above representation of $u^{\vecbold{\pi}}_{k}\paren{\vecbold{s}}$ in \eqref{Eqn: Stalling Probability} that shows the dependence of segment stalling (or, non-stalling to this effect) probability on receiver and playout curves explicitly will be helpful in our derivation to determine the structure of the optimum scheduling policy in the next section.   


\section{The Optimum Scheduling Policy}
\label{sec:optimum}
\begin{figure}[t] 
	  \centering
		\includegraphics[scale=0.5]{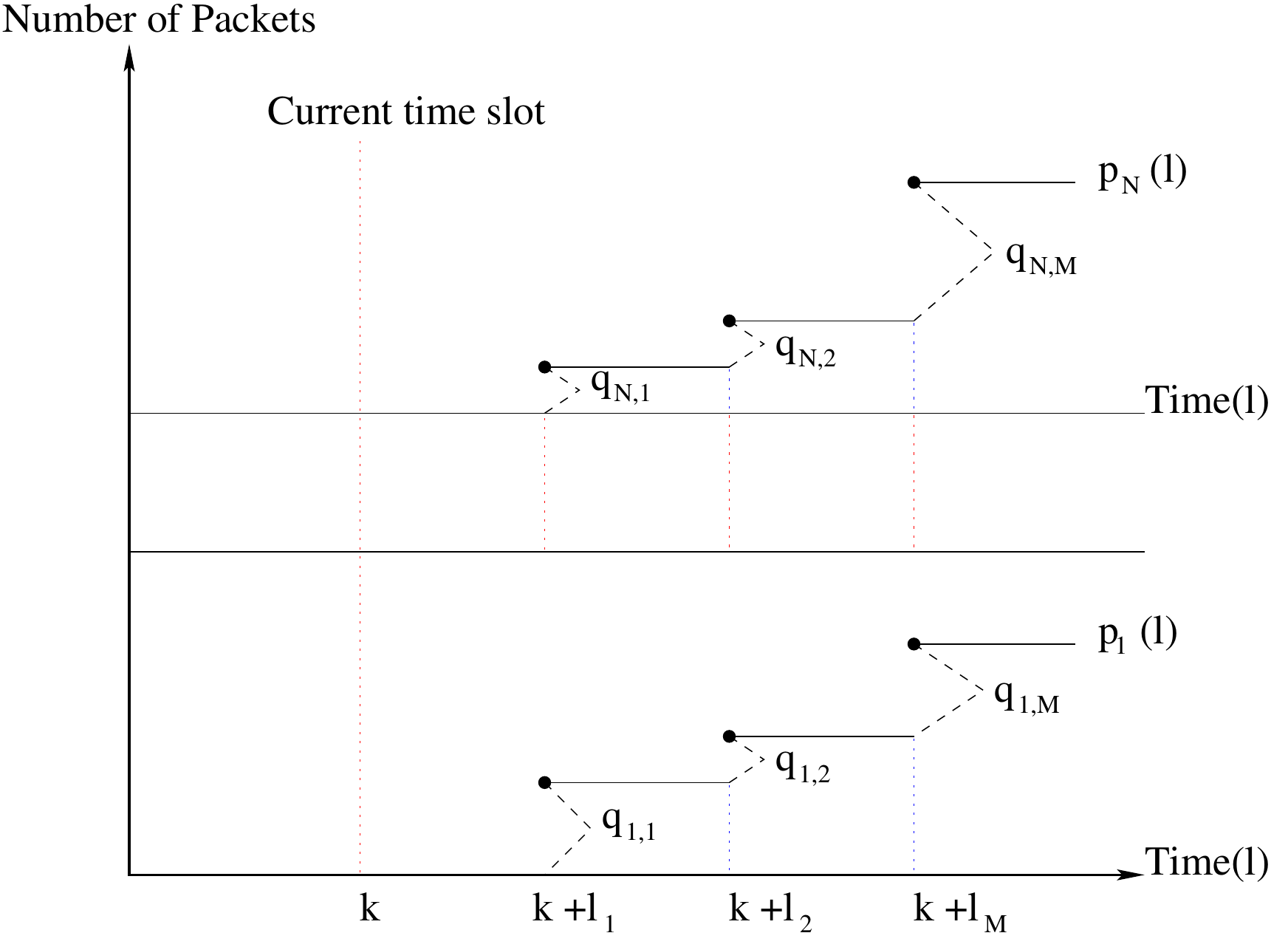}
		\caption{Schematic representation of playout curves at the beginning of time-slot $k$.}
	  \label{fig3}
\end{figure}
In this section, we derive the structure of the optimum scheduling policy  that solves the optimum scheduling problem introduced in Section \ref{Section: System Model} for maximizing the non-stalling event probability in multiuser video streaming systems.  In particular, it will be shown that a simple but practical greedy scheme that schedules users according to packet deadlines maximizes the segment non-stalling probability $u_k^{\vecbold{\pi}}\paren{\vecbold{s}}$ for all initial system states $\vecbold{s} \in \mathcal{S}$ as well as for time-slots $k \in \brparen{0,\ldots, T-1}$. We call this scheme the {\em blind deadline-based resource allocation} (BDRA) scheme.  Before we formally state the optimality of the BDRA scheme in Theorem \ref{Thm: Optimality of BDRA scheme}, which is the main analytical result of this paper, it would be helpful to explain the operational details of the BDRA scheme through a particular situation for facilitating the upcoming discussion and the exposition of the proof of its optimality.

To this end, consider the case where the current time-slot index is $k$ and assume that there are $M$ jumps in the playout curves of users at time-slots $k+l_{1},\ldots,k+l_{M}$, which is illustrated in Fig. \ref{fig3}.  These are the ordered time instants increasing from the smallest one to the biggest one with the last time instant $k+l_{M}$ coming no later than $T$.  We recall that such a jump occurring in the playout curve of a user corresponds to the additional data demanded by this user (in terms of number of packets) for smooth displaying of its video, and this data demand must be provisioned by the edge server in order to avoid video stalling at this user.  

We let $q_{i, m}$ denote the height of the jump at time-slot $k + l_{m}$ occurring at the playout curve of user $i$. Here, $q_{i, m}$ corresponds to the number of additional data packets requested by user $i$ between the deadlines $m-1$ and $m \leq M$. Therefore, we can consider the delivery of $q_{i, m}$ packets to user $i$ as a {\em task} with a {\em deadline} $k+l_{m}$. If this task is accomplished by the edge server for all deadlines, then no stalling event occurs at user $i$. The BDRA scheme simply prioritizes all such tasks based on their deadlines by instructing the edge server to conclude the tasks with the earliest deadlines first before proceeding to those with deadlines coming at later times.  If there are two or more users with the same deadline, the BDRA scheme can choose any one of such users without any loss of optimality.

\begin{theorem} \label{Thm: Optimality of BDRA scheme}
For given playout curves $\boldsymbol{p}$ and channel statistics $\boldsymbol{\beta}$, the BDRA scheme produces an optimal scheduling policy $\boldsymbol{\pi}^{bdra}$ i.e.,
\begin{equation}
u^{\boldsymbol{\pi}^{bdra}}_{k}(\boldsymbol{s})=u^{\star}_{k}(\boldsymbol{s})
\end{equation}
holds for all $k=0, \ldots,T-1$ and $\vecbold{s} \in \mathcal{S}$.
\end{theorem}
\begin{IEEEproof}
We will prove this theorem by induction. 

\emph{Base Case:}  We first consider the base case in which the optimum scheduling problem is solved for the last time-slot $T-1$.  If there are two or more deadlines in the beginning of time-slot $T-1$, no scheduling policy can achieve stalling-free video streaming for all users, and therefore all scheduling policies are the same in terms of their segment stalling probability performances in such cases.  On the other hand, if there is only one deadline in the beginning of time-slot $T-1$, the user associated with this deadline must be served to avoid a possible stalling event. This discussion shows that the BDRA scheme minimizes the segment stalling probability for the last time-slot.

\emph{Induction Step:}  Secondly, we consider a time-slot with index $k+1 \leq T-2$ and assume that $u^{\vecbold{\pi}^{bdra}}_{k+1}\paren{\vecbold{s}} = u^{\star}_{k+1}\paren{\vecbold{s}}$ for all $\vecbold{s} \in\mathcal{S}$.  Then, it is well-known from \cite{dynamicprog} that the optimal  decision for time-slot $k$ must satisfy the following condition
\begin{equation}\label{optconst}
d^{\star}_{k}\paren{\vecbold{s}_k} \in \argmax_{a \in \mathcal{A}}\brparen{\sum_{\boldsymbol{s} \in \mathcal{S}} P_{k}\paren{ \boldsymbol{s} \vert \boldsymbol{s}_k, a} u^{\star}_{k+1}\paren{\vecbold{s}}} 
\end{equation}
for all system states $\vecbold{s}_k \in \mathcal{S}$ in the beginning of time-slot $k$. 
Since the induction hypothesis asserts that $u^{\boldsymbol{\pi}^{bdra}}_{k+1}\paren{\boldsymbol{s}} = u^{\star}_{k+1}\paren{\boldsymbol{s}}$, (\ref{optconst}) can also be expressed as
\begin{equation}\label{optconst2}
d^{\star}_{k}\paren{\boldsymbol{s}_{k}}\in \argmax_{a\in A} F_{k}\paren{\boldsymbol{p},\left(a,\boldsymbol{\pi}_{k+1}^{bdra}\right),\boldsymbol{s_{k}}}.
\end{equation}

Note that the term $\paren{a,\boldsymbol{\pi}_{k+1}^{bdra}}$ in (\ref{optconst2}) is a tail scheduling policy that is obtained by concatenating an action $a$ and the tail policy $\boldsymbol{\pi}_{k+1}^{bdra}$. Next, we will show that $\vecbold{\pi}_k^{bdra} = \paren{d^\star_k, \vecbold{\pi}_{k+1}^{bdra}}$.  To this end, we will provide an alternative expression for $F_{k}\paren{\boldsymbol{p},\boldsymbol{\pi}_{k},\boldsymbol{s}_{k}}$ for any tail scheduling policy $\vecbold{\pi}_k$. 
Let there be $M$ deadlines at $k+l_{1},\ldots,k+l_{M}$ for a given playout curve $\boldsymbol{p}$ and system state $\vecbold{s}_k \in \mathcal{S}$ after the time-slot $k$, an example of which is illustrated in Fig. \ref{fig3}.  Let also the random variable $\lambda_{m}$ denote the first time-slot when all packets belonging to the first $m$ deadlines are delivered successfully.  We note that $\lambda_{m}$ depends on the tail scheduling policy $\boldsymbol{\pi}_k$ and $\vecbold{p}$, and $F_{k}(\boldsymbol{p},\boldsymbol{\pi}_{k},\boldsymbol{s}_{k})$ can be expressed in terms of $\left\{\lambda_{m}\right\}^{M}_{m=1}$ as
\begin{equation}\label{sfvsp}
F_{k}(\boldsymbol{p},\boldsymbol{\pi}_{k},\boldsymbol{s}_{k})=\PRP{\bigcap^{M}_{m=1} \brparen{\lambda_{m}\leq k+l_{m}} \Big \vert \ \vecbold{s}_k}. 
\end{equation}

Consider now the random variable $\tau_{m}$, which denotes the total number of time-slots required to send  all $\sum^{N}_{i=1}q_{i,m}$ packets associated with the deadline at $k+l_{m}$ successfully.  Under the BDRA scheme, the relationship between $\lambda_{m}$ and $\brparen{\tau_{i}}_{i=1}^m$ is $\lambda_{m}=k+\sum^{m}_{i=1}\tau_{i}$.  Hence, using \eqref{sfvsp}, we obtain
\begin{equation}\label{sfvsp2}
F_{k}\paren{\boldsymbol{p},\boldsymbol{\pi}^{bdra}_{k},\boldsymbol{s}_{k}}=\PRP{ \bigcap^{M}_{m=1} \brparen{\sum^{m}_{i=1}\tau_{i} \leq l_{m}} \Big \vert \ \vecbold{s}_k}.
\end{equation}


Assume now that we choose an action $a \neq d_k^{bdra}\paren{\vecbold{s}_k}$ and form a tail scheduling policy $\paren{a, \vecbold{\pi}_k^{bdra}}$. For this tail  scheduling policy, we will show that $F_k\paren{\vecbold{p}, \paren{a, \vecbold{\pi}_{k+1}^{bdra}}, \vecbold{s}_k} \leq F_{k}\paren{\boldsymbol{p},\boldsymbol{\pi}^{bdra}_{k},\boldsymbol{s}_{k}}$.  Let the scheduled user $a$ has the first deadline at $k+l_{j}$ for some $j \geq 2$.  Since time-slot $k$ is allocated for user $a$, and the slot allocation is done according to the tail policy $\boldsymbol{\pi}^{bdra}_{k+1}$ in the remaining time slots, we can write $F_k\paren{\vecbold{p}, \paren{a, \vecbold{\pi}_{k+1}^{bdra}}, \vecbold{s}_k}$ as 
\begin{eqnarray}\label{sfvsp3}
\lefteqn{F_k\paren{\vecbold{p}, \paren{a, \vecbold{\pi}_{k+1}^{bdra}}, \vecbold{s}_k}} \hspace{8cm}\nonumber\\
\lefteqn{= \PRP{\bigcap_{m=1}^M \brparen{\sum_{i=1}^m \tau_i \leq l_m - \I{m < j} } \Big \vert \ \vecbold{s}_k},} \hspace{6.5cm}
\end{eqnarray}
where $\I{m<j}$ is an indicator function that returns $1$ if the inequality $m<j$ holds.\footnote{The random variables appearing in \eqref{sfvsp2} and \eqref{sfvsp3} must be considered to be equal in distribution.} Comparing \eqref{sfvsp2} and \eqref{sfvsp3}, we conclude that $F_{k}\paren{\boldsymbol{p},\boldsymbol{\pi}^{bdra}_{k},\boldsymbol{s}_{k}} \geq F_k\paren{\vecbold{p}, \paren{a, \vecbold{\pi}_{k+1}^{bdra}}, \vecbold{s}_k}$ for any $a \in \mathcal{A}$. This result implies that $\vecbold{\pi}_k^{bdra}$ is the optimum tail scheduling policy starting from any time-slot $k$ onwards, and hence $\vecbold{\pi}^{bdra}$ is the solution of the optimum scheduling problem given by \eqref{Eqn: Scheduling Optimality}.    
\end{IEEEproof}

An important corollary of Theorem \ref{Thm: Optimality of BDRA scheme} is that the optimum scheduling policy minimizing the segment stalling probability does not depend on the statistical knowledge $\vecbold{\beta} = \paren{\beta_1, \ldots, \beta_N}$ of the wireless channel between the edge server and the users. This observation may seem counter-intuitive at a first glance.  In particular, it can be conjectured that we should always perform better if we take channel statistics into account while giving scheduling decisions in each time-slot. However, the particular solution constructed for the optimum scheduling problem in Theorem \ref{Thm: Optimality of BDRA scheme}, i.e., the BDRA scheme, shows that we cannot improve the segment stalling probability even if we utilize the statistical channel knowledge.\footnote{The solution for the optimum scheduling problem in \eqref{Eqn: Scheduling Optimality} is not necessarily unique, and there may exist other resource allocation policies utilizing wireless channel statistics and achieving the same performance with the BDRA scheme. The determination of the complete solution set for \eqref{Eqn: Scheduling Optimality} is outside the scope of the current paper.} The point here is that the dynamic playout curve updating procedure embedded in the BDRA scheme already includes the effect of the packet drop probabilities of the users, and this is sufficient to make the BDRA scheme an optimum scheduling policy for multiuser video streaming systems.

This observation has some important practical ramifications.  Firstly, implementation of the BDRA scheme avoids any channel estimation issues to learn channel conditions before it starts its operation. In particular, implementation of a channel estimation algorithm suited for the particular requirements of video streaming coupled with an efficient and high-throughput feedback protocol design (for frequency-division-duplexing systems) from users to the edge server may become an onerous task for delay sensitive video traffic.  

Secondly, perhaps the most importantly, the BDRA scheme can be implemented as an {\em add-on} solution to the existing video streaming systems, especially to the DASH based systems, for improving their efficiency.  Therefore, it must be backward-compatible with them for all practical purposes, rather than necessitating a substantial re-design of a video streaming system. Besides improving the  efficiency of video streaming systems by minimizing the stalling event probability, its simple and channel statistics invariant nature makes the BDRA scheme an ideal backward-compatible solution for serving this purpose.  Finally, the BDRA scheme has only polynomial-time computational complexity due to ordering users according to corresponding deadlines, and hence easy to execute in real-time.  In the next section, we present a particular NS-3 implementation of the BDRA scheme integrated into a DASH based video streaming system to illustrate its aforementioned benefits.


\section{Implementation and Design Issues}
\label{sec:Implementation}
\subsection{Implementation}

 Another important corollary of Theorem \ref{Thm: Optimality of BDRA scheme} is that the optimum scheduling policy, minimizing the segment stalling probability, allocates the time slots to the current client until  all the packets in the corresponding GoP are sent.  This is because, the BDRA scheme allocates the time slots to the clients in the order of upcoming deadlines, and all the packets belonging to the same GoP has the same deadline.  An important practical consequence of this fact is that BDRA scheme can be implemented at the application layer completely oblivious of the operation of the lower layer protocols.  The only information required by BDRA scheme when implemented at the application layer is the acknowledgment of completion of GoP, which can be effectively inferred when the client sends a new HTTP-GET message for the subsequent GoP. We note that GoP based video transmission methods are already known in the literature \cite{GoPbased}. However, in this work we show that by utilizing certain features of DASH structure an optimal GoP based policy can be constructed without using an additional feedback mechanism between the server and clients.

The operation of DASH based video streaming can be further conceptualized as follows. The client begins the streaming period by first requesting the associated MDP file. The edge server acts as a {\em web proxy} for the client, requesting the MDP file from the video content delivery server on its behalf.  A copy of the received MDP file is stored at the edge server, whereas another copy is forwarded to the client. Based on the received MDP file and the estimated network throughput, the client requests the first segment among all available representations with HTTP-GET command and the video streaming from the main server starts. At the same time, the edge server observes the HTTP-GET command for the first segment and defines the deadline of the user according to initial buffer duration. The received files from the main server stored in the edge server in a sequence of GoPs via utilizing the MDP file. The edge server controls the deadlines of the existing users and executes the BDRA scheme. Whenever a client receives a GoP file successfully, it sends a HTTP-GET \footnote{The smallest size of the sub-segment requested can be equal to the size of one GoP and it is requested via HTTP partial GET command.} command for the next GoP and this command is not conveyed to the main server but utilized by the edge server to update the deadline of the client i.e., the deadline of the client is extended by the GoP duration. If the deadline of the client expires, the edge server senses that a stalling event has occurred and extends the deadline of the client by the rebuffer duration. Since the HTTP-GET commands are utilized to update deadlines, the BDRA scheme executed in the edge server does not need to trace the client buffer constantly, which is critical to reduce the feedback load between the client and the edge server. Note that the download times of each GoP, and thus, the arrival time of the next HTTP-GET command may vary with respect to the size of the GoP and the conditions of the channel between the edge server and the client.
\subsection{Design Issues}
There are three important design issues for integrating the BDRA scheme with the existing protocol stack and DASH protocol. We discuss them below, starting with the issue of secure HTTP connection requests from the clients.    
\subsubsection{BDRA Scheme with Secure Video Streaming}
HTTP secure (HTTPS) is a newly emerging variant of the HTTP protocol in order to offer increased levels of privacy and security to end users on-demand \cite{Secure1}, \cite{Secure2}, \cite{Secure3}. In particular, video streaming services such as YouTube and Netflix already provide the secure end-to-end connection option through HTTPS connection. The secure connection in HTTPS is established through an authentication process in which a third party certification authority ensures the authenticity of the presented certificate by the streaming server \cite{Secure2}. One challenge that arises with HTTPS connection to implement the BDRA scheme at the access point is that only servers with a certain certificate can observe user video demands and corresponding GoP statistics \cite{Secure3}. This implies that the access point, if not equipped with the correct certificate, cannot observe the GoP deadlines for prioritizing the scheduling decisions. A potential resolution of this problem with HTTPS connections for video streaming is to notify the BDRA execution point with deadline information of the streaming video files without compromising the security of the content of these files. This can be achieved by assigning a unique identifier to each GoP and its corresponding deadline, and then relaying this information from the main server to the access point. The resource allocation mechanism at the access point utilizes only this information to allocate resources over multiple end users requesting different video files over shared communication resources.              
\subsubsection{BDRA Scheme with GoP Size Adaptation and Fairness}
The main promise of the derived BDRA scheme is its ability to minimize system-wide stalling event probability. This property of the BDRA scheme holds correct independent of GoP sizes of multiple users streaming the video files simultaneously. In addition, the simple structure of the BDRA scheme can be further utilized to improve end user experience with widely varying GoP sizes. In particular, the resource allocation mechanism at the access point can minimize the total number of streaming interruptions aggregated over all end users by considering GoP sizes of the files having the same deadline. For example, one can envision a scenario in which there are three users with the same deadline having GoP sizes of $8$, $5$ and $3$ packets, and the access point is able to send $8$ packets up to the corresponding deadline. Hence, the overall system experiences a stalling event at this deadline since there will be at least one end user whose packets cannot be delivered on time. However, if we start allocating time slots to the users in the order of increasing GoP sizes, then only the user with the largest GoP size will experience a service interruption, whereas there will be two of them experiencing such an interruption in the reverse order. 

In our implementation of the BDRA scheme for obtaining its performance figures, we assumed that the BDRA scheme assigns a priority to users according increasing GoP sizes if there are multiple users with the same deadline.\footnote{ We note that if the channel rates are known, then instead of ordering clients according to GoP sizes, clients with the same deadline can be ordered  according to GoP transmission time.} Another variation of the BDRA implementation one can consider here is to choose a user randomly when they have the same deadline to achieve a degree of fairness among them. The probability distribution for prioritizing the end users can even depend on their respective GoP sizes to strike a balance between fairness and stalling performance. As this discussion makes it clear, the simple structure we obtained for the optimum slot-based resource allocation lends itself to various modifications, and performance enhancements depending on the objectives to optimize are possible. Since the computing capabilities of edge servers are increasing rapidly, the more complex versions of the BDRA scheme  than the one considered in this paper can be implemented at the edge server with new product roll-outs.

One final remark we would like to make here is about the GoP durations of the video files being streamed.  When we use the term ``GoP duration" in the paper, we refer to the time duration measured in terms of the ratio of GoP sizes to the video frame rates. To the best of our knowledge, the on-demand video streaming services keep the GoP duration fixed in practical implementations. For instance, YouTube recommends GoP-Size-to-Frame-Ratio to be $0.5$. Hence, our experiments in the paper assume an identical GoP duration for the video files being streamed, which implies that an increase in the GoP size of a video file results in a corresponding increase in the video frame rate without a change in the GoP duration. 
  
\begin{figure*}
    \centering
         \begin{subfigure}[b]{0.3\textwidth}
        \includegraphics[scale=0.3]{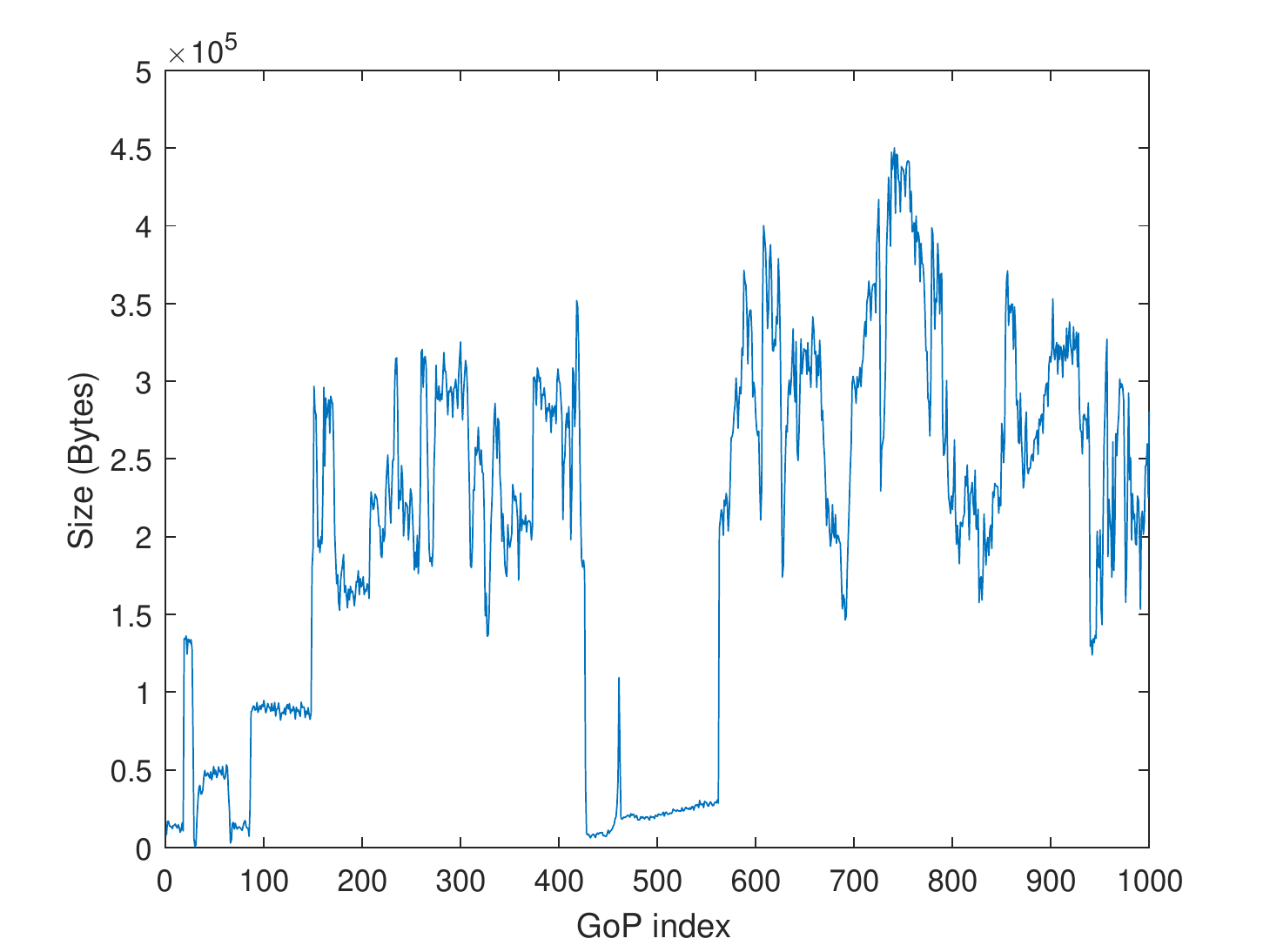}
        \caption{Tokyo Olympics}
    \end{subfigure}
    \begin{subfigure}[b]{0.3\textwidth}
        \includegraphics[scale=0.3]{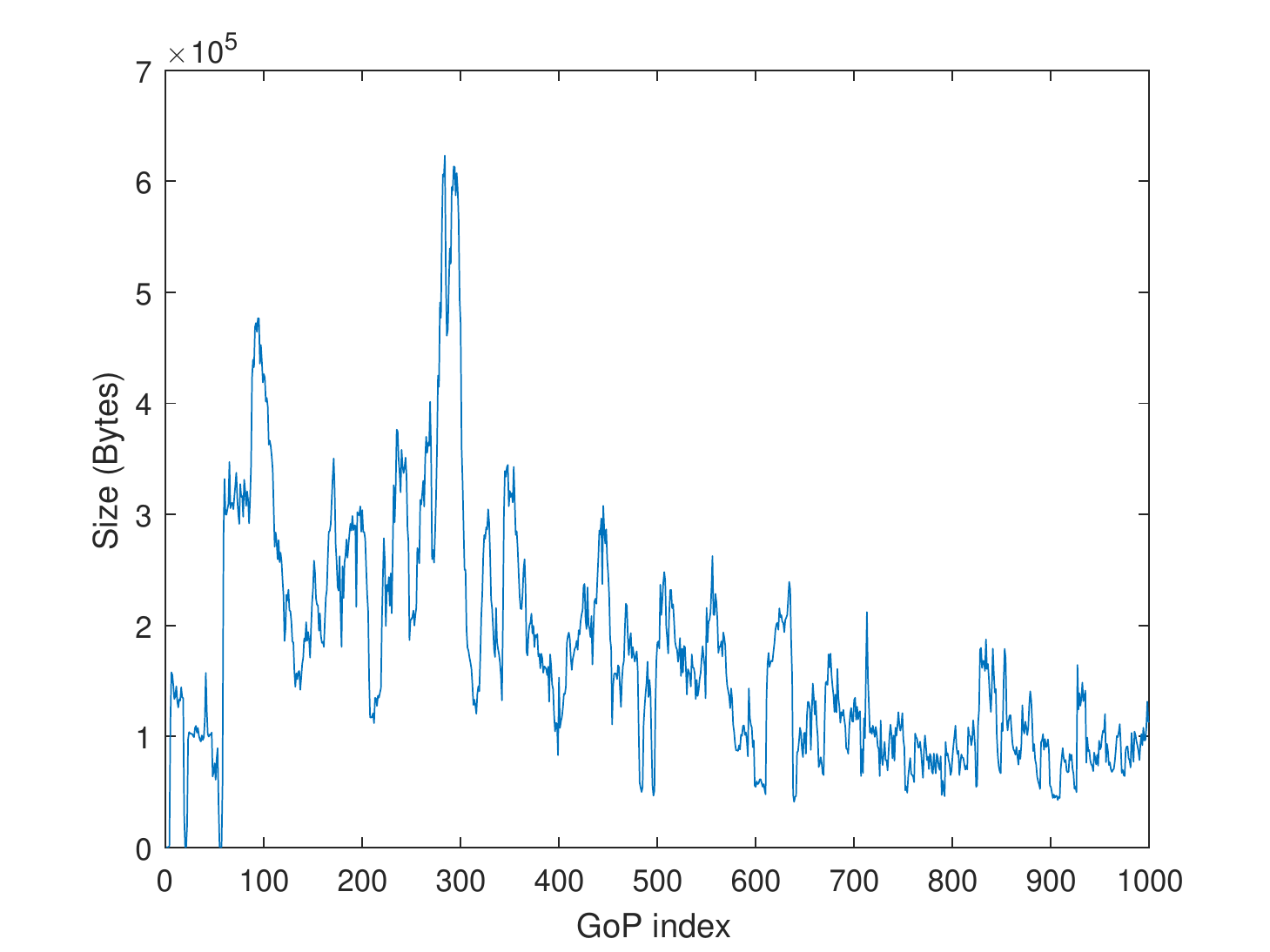}
        \caption{Silence of the Lambs}
        \end{subfigure}
         \begin{subfigure}[b]{0.3\textwidth}
        \includegraphics[scale=0.3]{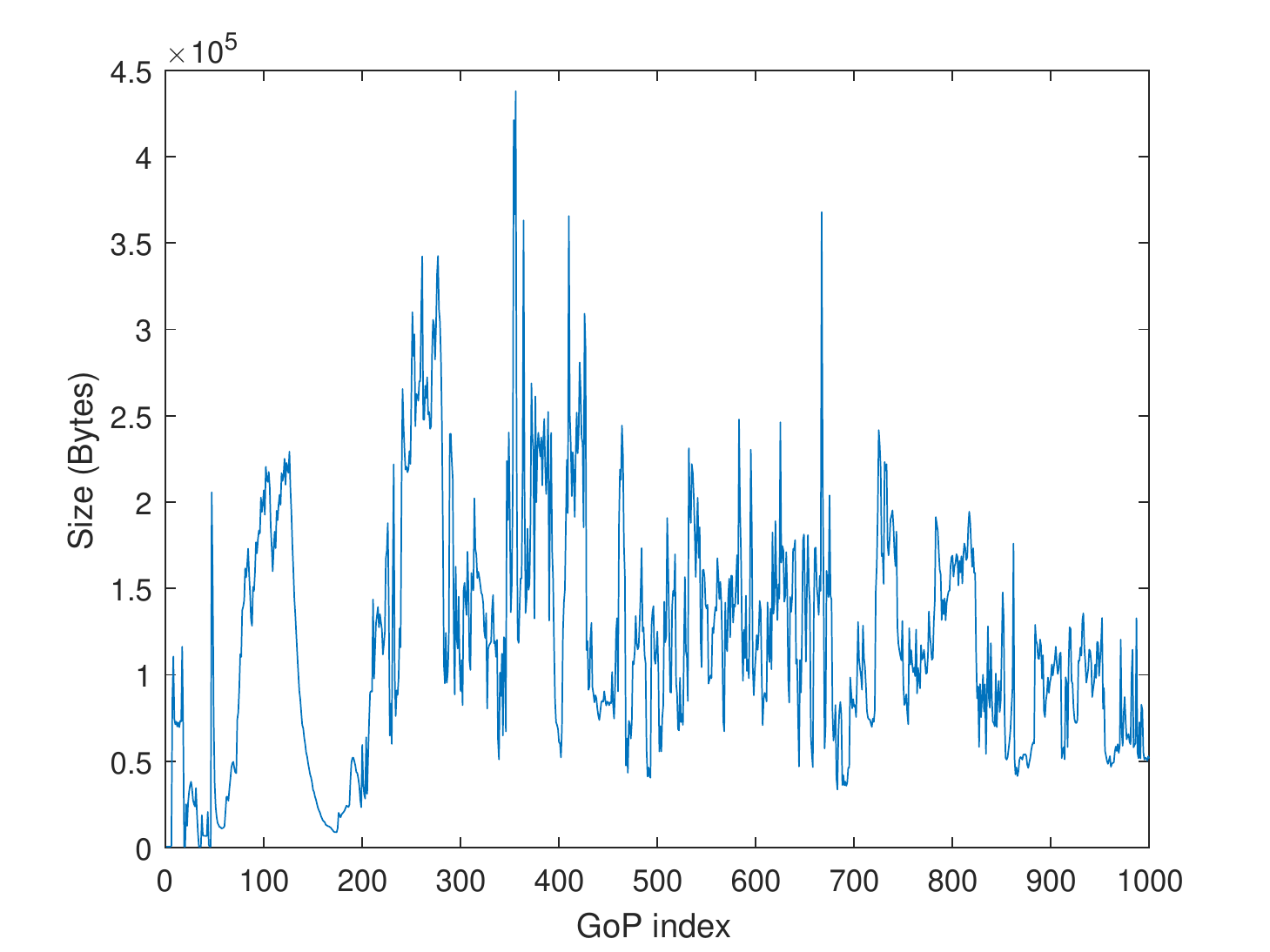}
        \caption{Star Wars, $QP=10$}
    \end{subfigure}
    \begin{subfigure}[b]{0.3\textwidth}
        \includegraphics[scale=0.3]{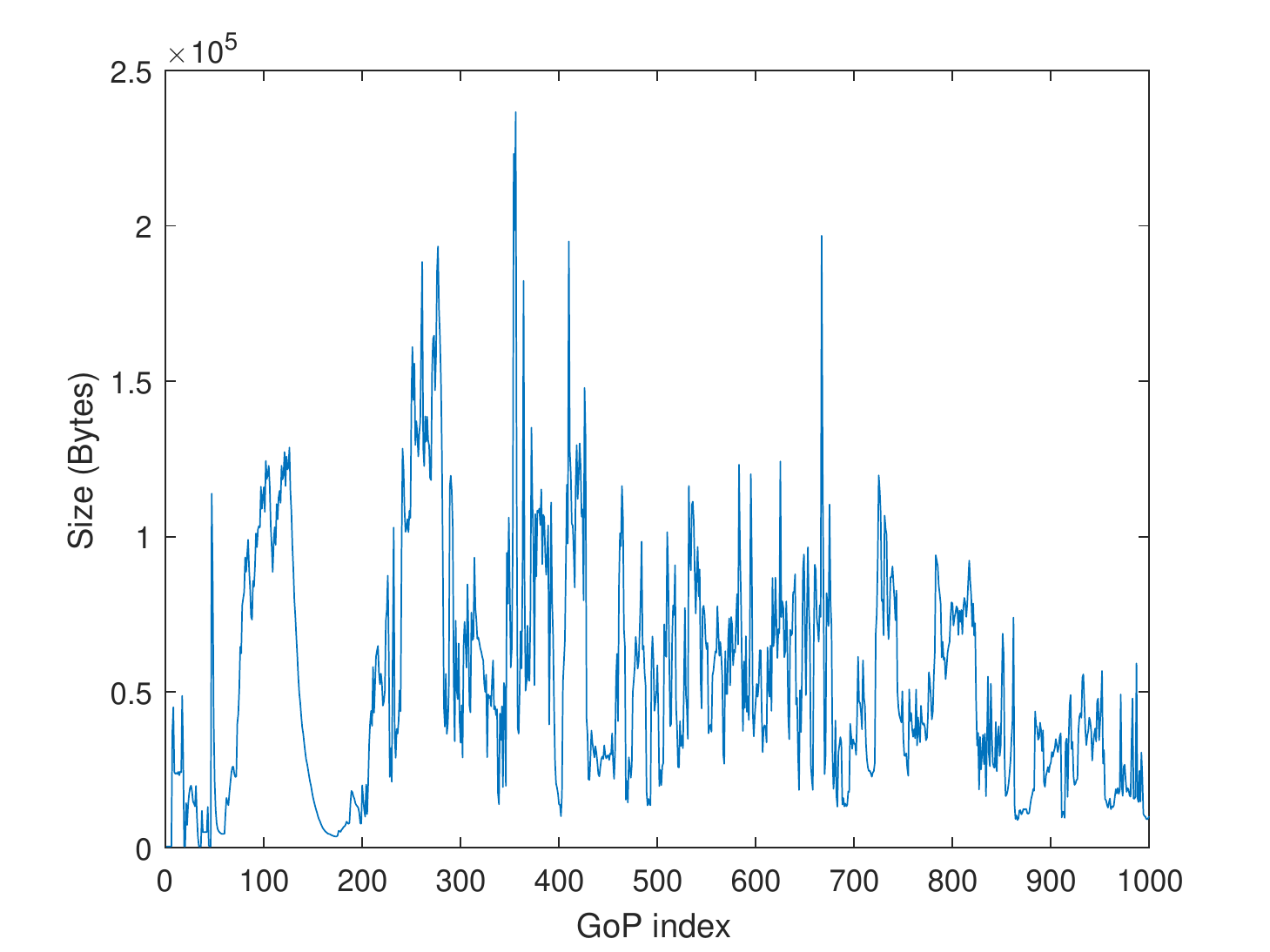}
        \caption{Star Wars, $QP=16$}
    \end{subfigure}
         \begin{subfigure}[b]{0.3\textwidth}
        \includegraphics[scale=0.3]{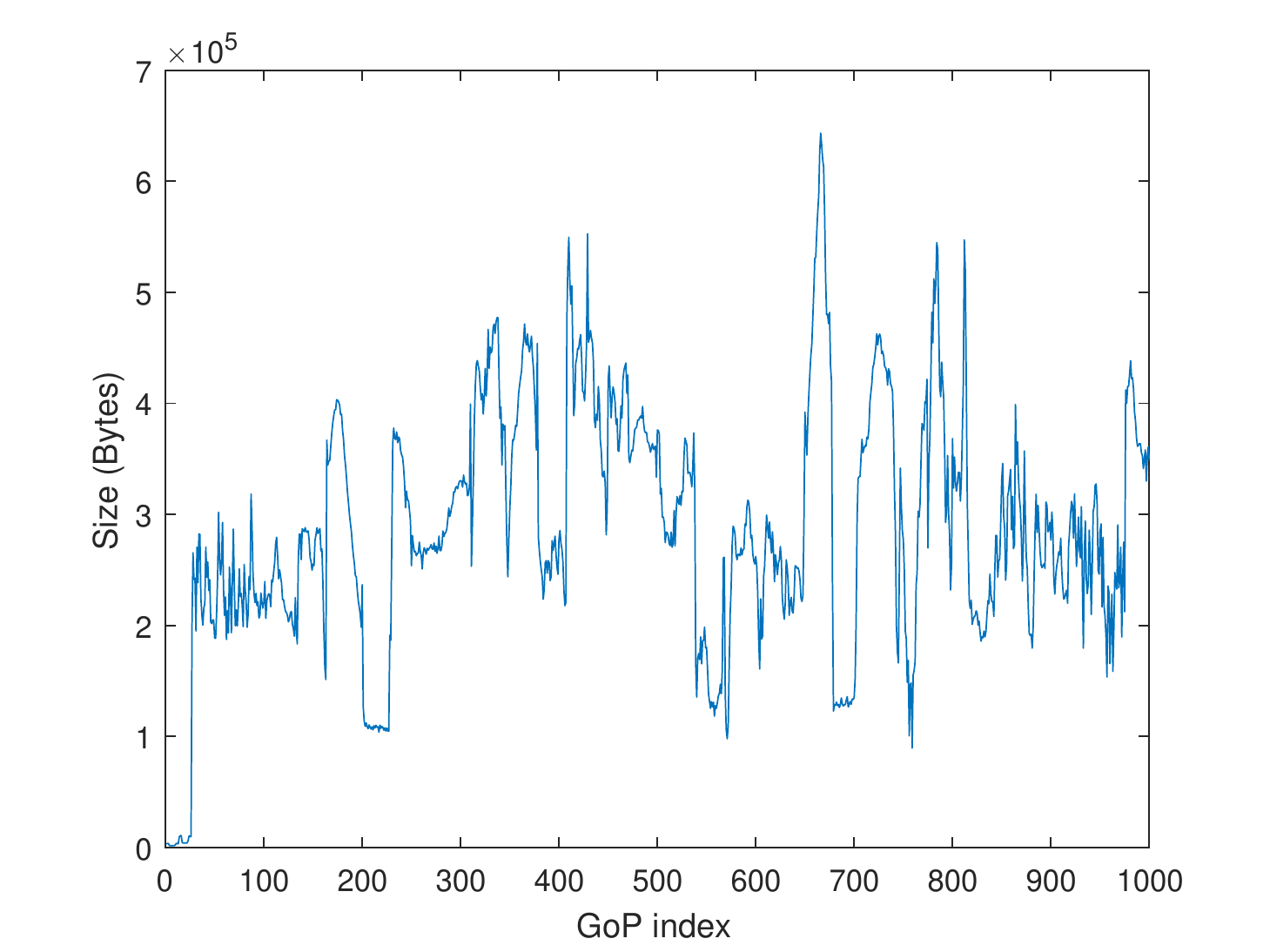}
        \caption{NBC News}
    \end{subfigure}
    \begin{subfigure}[b]{0.3\textwidth}
        \includegraphics[scale=0.3]{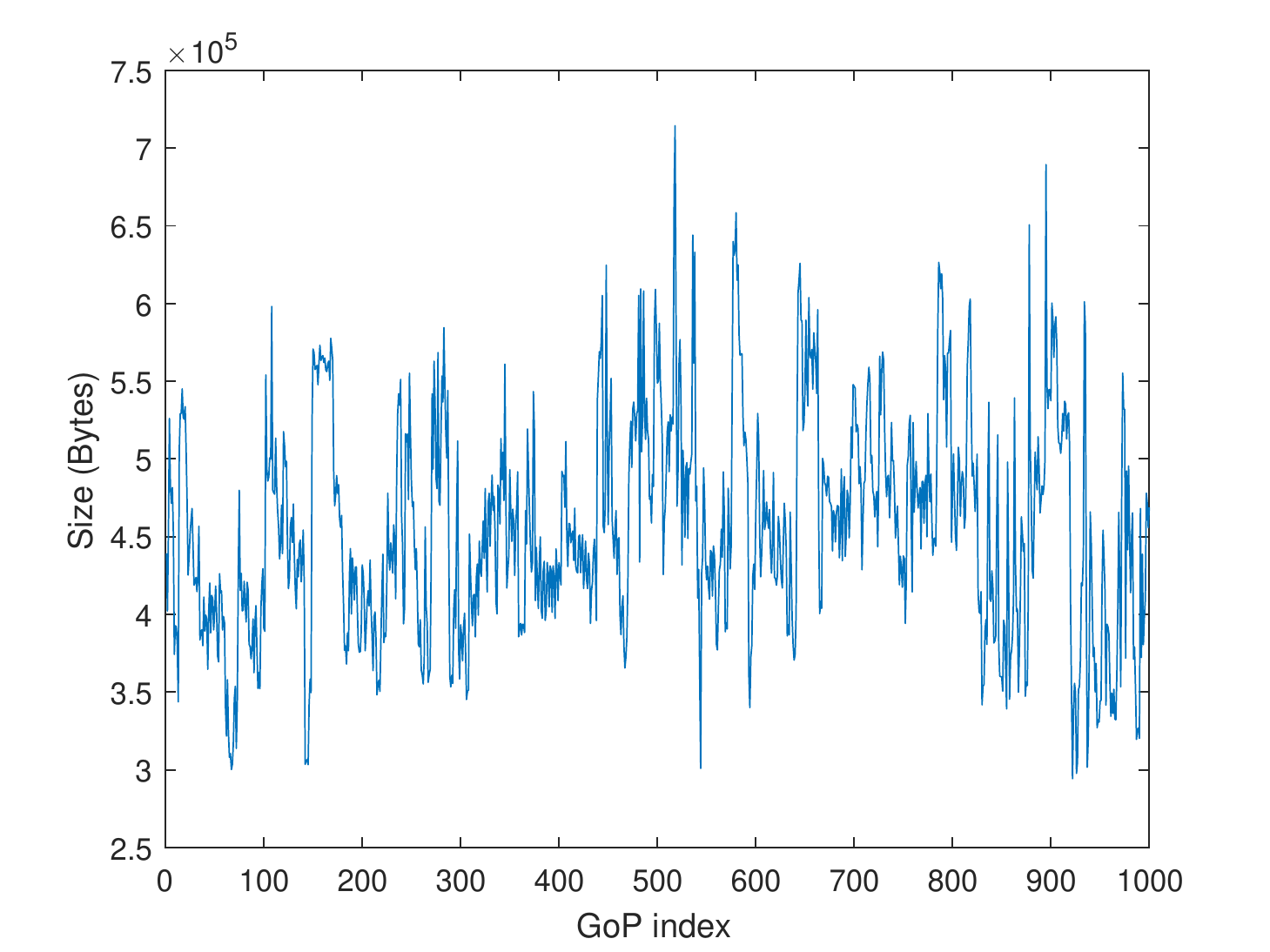}
        \caption{Sony Demo}
    \end{subfigure}
    \caption{Variation of GoP size over time.}\label{fluc}
\end{figure*}        
\subsubsection{ BDRA Scheme  with Redundant Chunk and Multiple Segment Requests} 
 An important adaptive feature of the existing DASH implementations is the redundant requests \cite{redundant}. In particular, if an end user senses that the network is lightly loaded (via bandwidth estimation), it may request a higher quality version of GoP/segment that is already buffered but not played. These redundant chunk requests are responded in a best effort way so that there is no guarantee that the user receives the higher quality version before starting to play the buffered version.

In our implementation of the BDRA scheme for simulations, we do not consider such redundant GoP requests. More specifically, these redundant chunk requests can be easily detected and discarded at the edge server since they will point to a deadline which has already been served. The reason for selecting this BDRA  design is to identify the effects of network load on the frequency of stalling events experienced by the end users. However, similar to the case of GoP size adaptation and fairness above, a modified version of the BDRA scheme that does not discard redundant chunk requests can also be implemented by prioritizing such requests after serving the end users with the current deadline according to the BDRA rule. In these implementations, the end users will experience more frequent streaming interruptions at the expense of having higher streaming video quality due to elevated levels of network load.

Another notable design issue regarding the existing implementations of the DASH protocol is the ability of users to request multiple segments/chunks by means of a single range-request.  There can be around $20$ GoPs in a single range-request. The structure of the derived BDRA scheme also exhibits agility against such multiple GoP requests from an implementation point of view. In particular, the sole purpose of sending a HTTP-GET request for each GoP in the derived BDRA implementation is to inform the access point about the successful delivery of the requested GoP so that the deadlines of the corresponding user can be updated accordingly. With multiple GoP requests, the access point will need to wait until the next such request before updating the deadlines of the corresponding user. To put it another way, multiple GoP requests transform the notion of ``GoP" in the derived BDRA  implementation into a notion of ``super-GoP", and the reception of a super-GoP request triggers the access point to update the deadlines for the subsequent GoPs of the streamed video file. 

An important remark here is the possibility of such super-GoP requests giving rise to a deterioration in the performance of the BDRA scheme to minimize the frequency of stalling events experienced by the users. The introduced BDRA implementation in this paper depends on a HTTP-GET request for each GoP, which requires a minimal modification at the client side, with a substantially improved video streaming experience in terms of the number of service interruptions. This performance boost is not available without the obtained BDRA-DASH integration, and hence super-GoP requests are beneficial in such a setting from the perspective of minimizing communication overhead between the users and the content distribution servers. However, with an BDRA scheme implementation integrated into the DASH protocol, it is an extra design problem to determine whether or not super-GoPs are still beneficial and if they are so, to decide about the number of GoPs to be included in each super-GoP request.  Last but not least, we can always consider other more demanding but useful alternative implementations of the BDRA scheme in order to accommodate super-GoP requests such as having an BDRA-assistant link layer control mechanism for conveying the GoP acknowledgment messages to update the deadlines at the access point.



\section{Numerical Results}
\label{sec:numerical}

In this section, we demonstrate the performance of the BDRA scheme as compared to other blind resource allocation schemes under realistic channel and network conditions.  All simulations are performed in NS-3 simulation environment.  By virtue of our proof in Section \ref{sec:optimum}, we know that the BDRA scheme is the optimum algorithm in order to minimize the frequency of video streaming interruptions for cases in which only the statistical knowledge of channels is available at the server-side. From this perspective, our main intention with NS-3 simulations in this part of the paper is to illustrate the {\em optimality gap} between the BDRA algorithm and other selected rate-fair resource allocation schemes. As a result, with this intention in the paper, we only compare the performance of the BDRA algorithm with other potential resource allocation mechanisms whose operation does not require knowledge about either network throughput rates, or channel quality indicators, or detailed client operation as different from most existing work in the literature \cite{proxy1, proxy2, channelbased, Colonnese15, networkassist2, sdn1, sdn2, bufferaware, networkassist}.

Recall that our protocol and its subsequent analysis is oblivious to the operation of lower layer networking stacks, but considers only whether the video packets are delivered to the end-user by their deadlines or not.  An important question arises on how the performance of this application layer protocol is affected by the operation of the lower layer protocols, i.e., specifically TCP congestion control protocol, and under general channel loss models.  Hence, in our simulations, we first considered a general Markov modulated channel model with packet loss varying among the states.  We also considered both an ideal cross-layer mechanism, which provides perfect and instantaneous feedback to our application layer protocol, and a realistic TCP protocol that performs retransmissions and adjusts the congestion window size based on packet losses.

We consider two different types of experimental setups. The goal of the first set of experiments is to verify the predictions of our theoretical results in Section IV by focusing on small time intervals (i.e., $20$ GoPs corresponding approximately to $10$ seconds). This first set of experiments are repeated $1000$ times with different NS-3 seeds, which corresponds to a long time interval of approximately $2.8$ hours in the ergodic limit sense.  In the second type of experiments, on the other hand, we consider various video files with the number of GoPs ranging from $480$ (i.e., corresponding  approximately to $4$ minutes) to $1200$  (i.e., corresponding approximately to $10$ minutes).  Our experiments indicate that the video duration does not have an impact on the performance of the BDRA scheme. Hence, considering the video durations in on-demand streaming services such as YouTube as well as the observation of video duration having minimal effect on the performance of the BDRA scheme , video file durations ranging from $4$ to $10$ minutes  provide substantive evidence for the performance improvements to be gained through the BDRA scheme in DASH based video streaming services.  We relegate the implementation of a prototype platform with real clients dynamically joining to and leaving the system over longer time horizons on the order of weeks to a future study. As a final note, although  it can be easily implemented along with the BDRA scheme,  we do not consider the client-side quality selection mechanism for subsequent video segments in the simulations until subsection \ref{sec: adaptive} . That is,  all subsequent segments (and sub-segments) are of the same quality in our simulations until subsection \ref{sec: adaptive}.  This allows us to more clearly demonstrate the improvement in the segment stalling probability provided by the derived BDRA scheme.
\begin{figure*}
    \centering
         \begin{subfigure}[b]{0.45\textwidth}
        \includegraphics[scale=0.45]{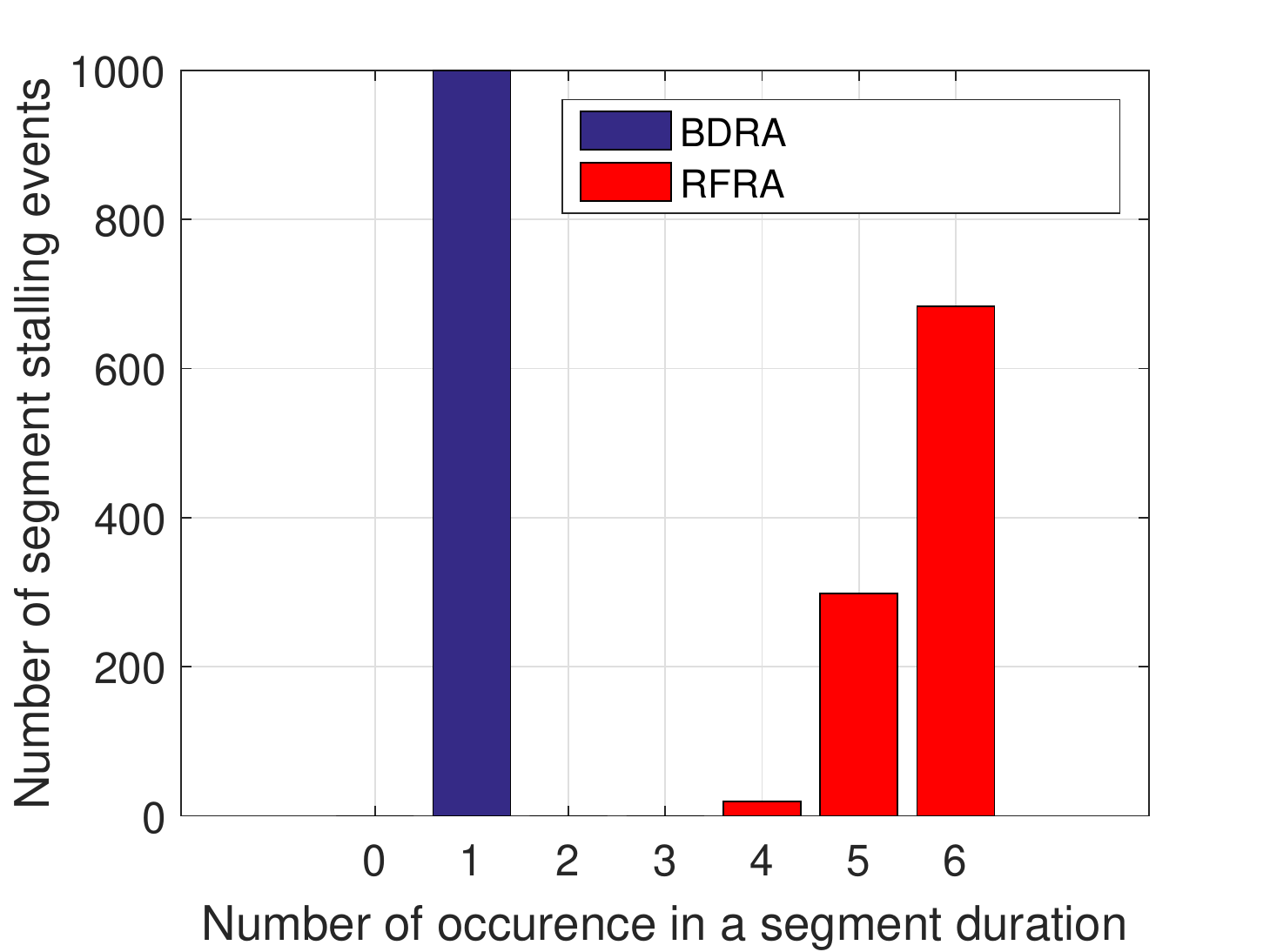}
        \caption{$\rho^{-1}= 1.3$.}
    \end{subfigure}
    \begin{subfigure}[b]{0.45\textwidth}
        \includegraphics[scale=0.45]{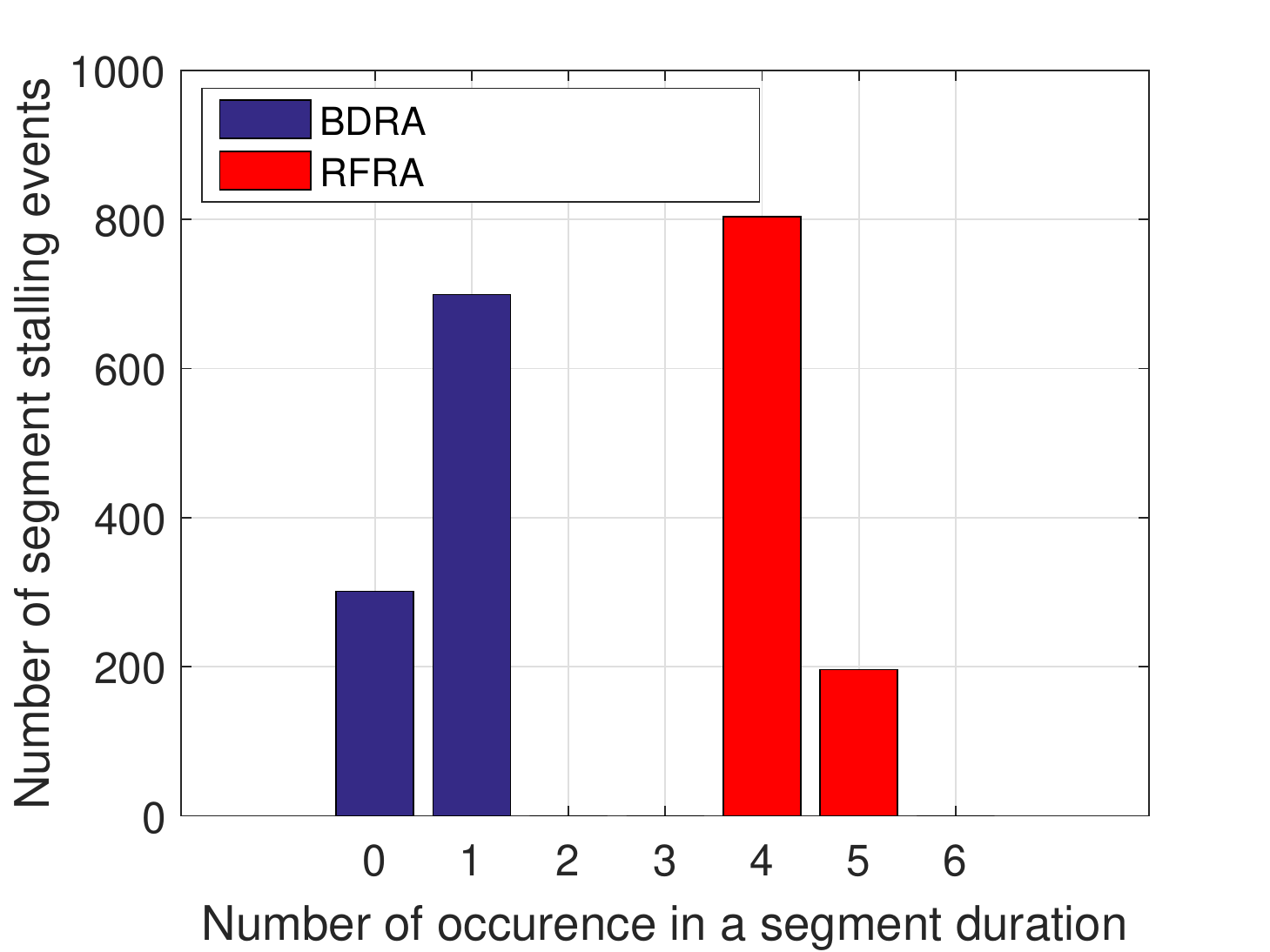}
        \caption{$\rho^{-1}=1.35$}
        \end{subfigure}
    \caption{The number of segment stallings.  }\label{segstall}
    \end{figure*}
\subsection{Experimental Setup}
In the experiments, we use H.264/AVC video traces that are accessible on the internet \cite{trace1,trace2}. All video traces have CIF resolution (352 $\times$ 288) at $30$ frames per second, frame configuration of $1$ B frames in between I/P key pictures and GoP size of $16$ frames. The pool of videos considered in the simulations are named Tokyo Olympics, Silence of the Lambs, Star Wars IV, NBC News and Sony Demo. For each video file except Star Wars IV, we add video trace with quantization parameter (QP) of 10 and for Star Wars IV we add video traces with QP of 10 and 16.\footnote{A quantization parameter is used to determine the quantization level of transform coefficients in H.264/AVC. An increase of 1 unit in the quantization parameter means an increase of quantization step size by approximately 12 percent, which in turn means 12 percent reduction in the video-rate \cite{trace3}.} The segment size is assumed to be 10.66 seconds (i.e., 20 GoPs). 

 Due to the AVC encoding, although the GoP duration is fixed, GoPs in the same video file may have different sizes measured in terms of the number of bits contained in each GoP segment.  Variation in the GoP sizes over time is demonstrated for the video files used in our experiments in Fig \ref{fluc}. In parallel to GoP size variation, data requirement of the client also fluctuates over time. Note that one key advantage of the GoP based BDRA scheme is that it can  respond to the fluctuations in the data rate requirements from the end users. This adaptive feature of the BDRA scheme leads to more significant performance improvements over the rate-fair resource allocation schemes when high bit-rate video files are requested for streaming since the GoP size variation in high bit-rate video files is significantly higher than that in low bit-rate ones.\footnote{For instance, the variance of GoP size in Star Wars IV with QP 10 is almost three times larger than the variance of GoP size in Star Wars IV with QP 16, although the GoP size variation patterns are identical.}

The main motivation to use the CIF resolution format as opposed to using QCIF in this paper is the ease of accessing to CIF statistics for several video files through publicly available databases such as http://trace.eas.asu.edu. Since the main feature of the BDRA scheme is its ability to exploit the GoP size variation, its operation does not depend on the particular choice of the resolution format (i.e., CIF or QCIF) of the video signals. As long as our BDRA scheme is supplied with the GoP variation trends, it can exploit these variations in order to minimize the system-wide stalling event probability in a DASH-based video streaming platform. 

The data transmission channel between the edge server and a user is characterized by a data rate and the error model. For the error model, we use {\em RateErrorModel} class of NS-3. In the NS-3 environment, rate error model is implemented under the transport layer. Hence, TCP packets are dropped according to an underlying probability distribution. In the literature, packet error rate (PER) is considered to be in the range of $\left[10^{-4},10^{-2}\right]$ for the TCP simulations \cite{per}. In \cite{per2},  the authors analyze the relationship between the PER and quality-of-service by using the video traces  to calculate the peak signal-to-noise ratio (PSNR) of the received video files. These PSNR indicators are used to evaluate the mean opinion score (MOS). Their analysis reveals that  $\left[10^{-4},10^{-3}\right]$, $\left[10^{-3},3\times10^{-3}\right]$ and $\left[3\times10^{-3},10^{-2},\right]$ correspond to the quality-of-service levels  {\em excellent}, {\em good} and {\em satisfactory}, respectively. In order to conform with these existing results, we consider a  Markov modulated link model, where there are three states with packet drop probabilities $\left[0.001,0.002,0.005\right]$, respectively, with each state corresponding to a different level of quality-of-service. The state transition probability matrix $\mathbf{\Gamma}$ is taken as
\[
\mathbf{\Gamma}=
  \begin{bmatrix}
    0.3 & 0.6 & 0.1\\
    0.2 & 0.6 & 0.2 \\
        0.1 & 0.6 & 0.3
  \end{bmatrix}.
\]
 A state transition occurs  at every 0.5 seconds, and the packet loss probabilities remain constant in between state transitions.

 We note that although the derived BDRA scheme will achieve similar performance gains for different physical layer telecommunication technologies (i.e., its operation is independent of the particular physical layer implementation as long as the packet losses can be modeled in a probabilistic manner at the upper layers), this particular NS-3 implementation is closer to a $4$G/$5$G scenario in which all communication resources are allocated to the user with the earliest deadline.  In this scenario, the physical layer outage events due to fading at various time scales (e.g., fast and slow fading) will be observed as packet losses at higher network and transport layers.  These packet losses will, in turn, determine which video files to be sent based on the updated deadlines to minimize the probability of stalling event occurrences. Further, the derived Markov modulated link model captures the time-varying nature of wireless channels in this scenario.

Let $\lambda_{i}$ (packets/sec) be the average rate of video packets generated for user $i$, which is calculated as the ratio of total size of the requested video file and the duration of the video.  Also, let $r$ (bytes/sec) and $L$ (bytes/packet) be the fixed channel data rate  for successful transmissions and fixed packet size, respectively. Then, the {\em inverse}  utilization rate $\rho^{-1}$ is the ratio of the channel data rate and the cumulative video source rate, which is defined as,
\begin{equation}
\rho^{-1}=\frac{r}{L\sum^{n}_{i=1}\lambda_{i}}.
\end{equation}
Values of $\rho^{-1}$ close to $1$ correspond to a highly loaded network, whereas large values of $\rho^{-1}$  correspond to a lightly loaded network.  In the following, we only consider an {\em underloaded} network scenario in the sense of having $\rho^{-1} > 1$ since those are the cases in which stalling events can be avoided and the efficiency of a scheduling algorithm is more clearly observed.\footnote{We use inverse utilization rates in order to obtain a parametrization for describing how heavily loaded the network in question is through numbers larger than $1$ in our simulations below.} 

\subsection{Segment Stalling Probability Distribution}
In this subsection, we analyze the distribution of the stalling events per segment when the BDRA scheme and a rate-fair resource allocation (RFRA) scheme that divides available communication resources equally among the streaming clients are employed. We consider a network with six users with each one requesting a different video file, e.g.,  Tokyo Olympics with QP=10, Silence of the Lambs with QP=10, Star Wars IV with QP=10, NBC news with QP=10, Sony Demo with QP=10, and Star Wars IV with QP=16. The duration of the simulation is taken as 10.666 seconds, which is also the duration of a segment.  The experiment is repeated for 1000 times with different random seeds for the RateErrorModel class, which corresponds to a long time interval of approximately $2.8$ hours in the ergodic limit sense. 

For the purpose of discovering segment stalling probability distribution, we set the average rate of packet losses to $0.2$ for each user, which corresponds to the packet loss rate experienced at the link layer.  Whenever a packet is lost, we assume that there is a perfect and instantaneous feedback sent to the transmitter. Although high for a realistic experiment, the rationale behind fixing the packet loss rate at $0.2$ in this set of experiments is to increase the channel randomness to observe a wider spectrum of stalling events, and thereby to verify our analytical results. We note that the experiments demonstrating the system performance for longer video durations approximately ranging from $4$ minutes to $10$ minutes are conducted  according to the above Markovian channel model in the next subsection. The simulations are performed for $\rho^{-1}$ values of  $1.3$ and $1.35$.     
%
%

 The results are summarized as histogram plots of the number of stalling events per segment for each $\rho^{-1}$ value in Fig. \ref{segstall}. We first note that users experience a single stalling event per segment with very high probability when the  BDRA scheme is employed with $\rho^{-1} = 1.3$. However, when the RFRA scheme is used with the same inverse channel utilization rate, the users experience six stalling events per segment  approximately $70\%$ of the time and they never experience less than four stalling events per segment. Secondly, when $\rho^{-1}=1.35$, i.e., the network is more lightly loaded, the performance of BDRA scheme improves further as expected. In particular, with approximately $30\%$ of the time, the users experience no stalling events, and they experience only one stalling event per segment for the rest of the time. For the same case, although the performance of the RFRA scheme also improves, it is still far away from the BDRA scheme, with users experiencing four stalling events $80\%$ of the time and five stalling events  $20\%$ of the time. 
\begin{figure*}
    \centering
         \begin{subfigure}[b]{0.45\textwidth}
        \includegraphics[scale=0.45]{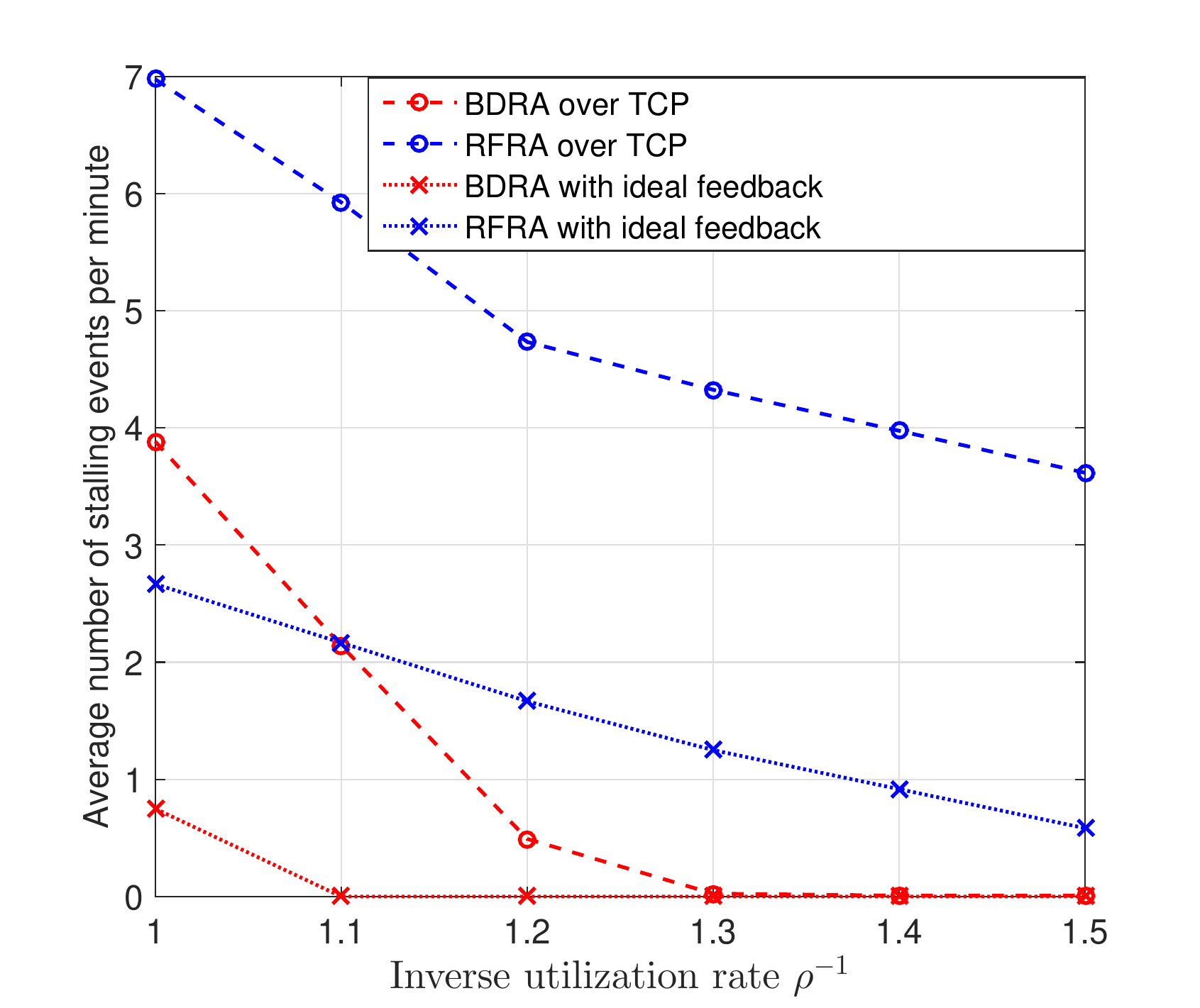}
        \caption{Video duration is $4\times16/15$ minutes.}
    \end{subfigure}
    \begin{subfigure}[b]{0.45\textwidth}
        \includegraphics[scale=0.45]{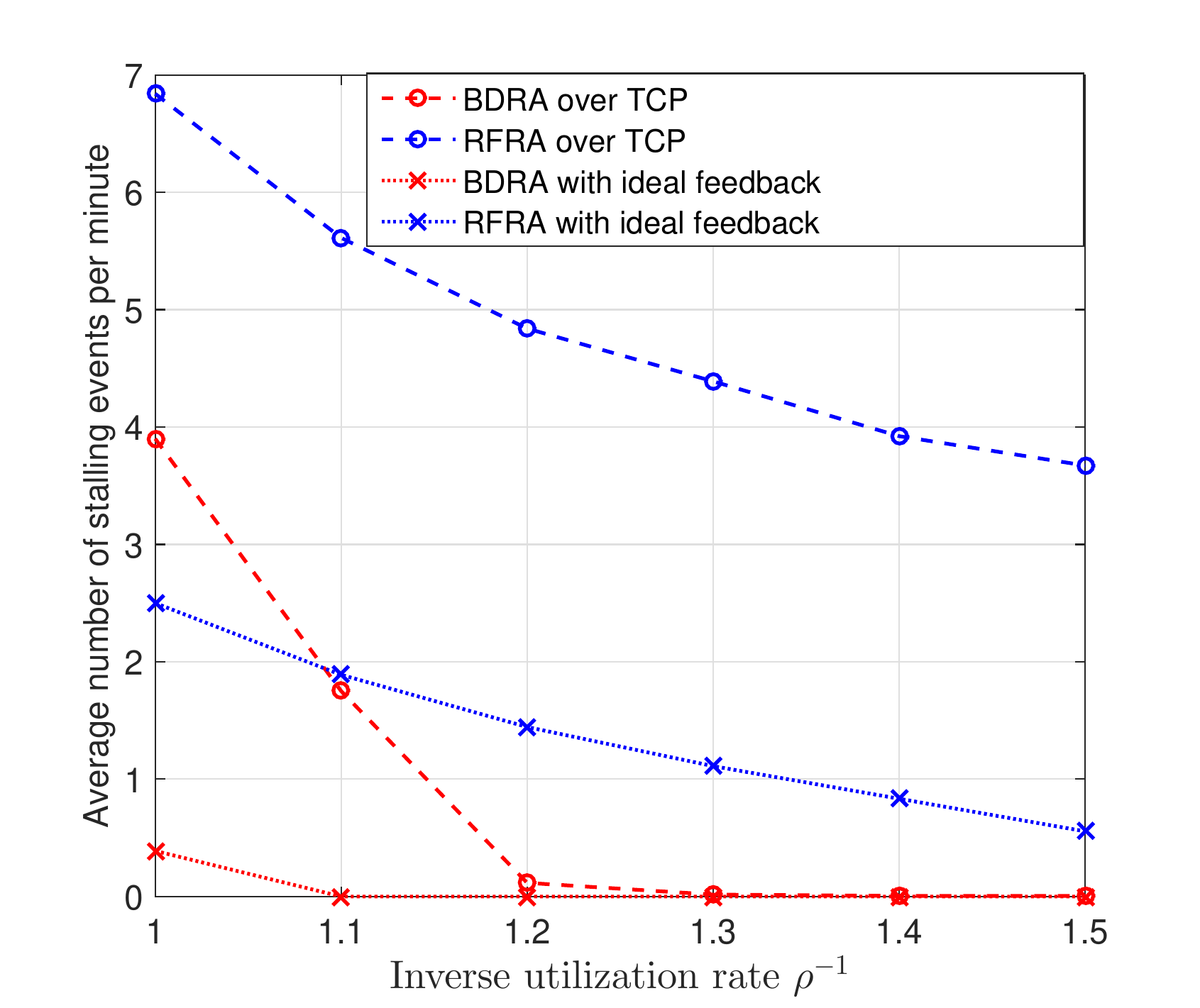}
        \caption{Video duration is $6\times16/15$ minutes.}
        \end{subfigure}
         \begin{subfigure}[b]{0.45\textwidth}
        \includegraphics[scale=0.45]{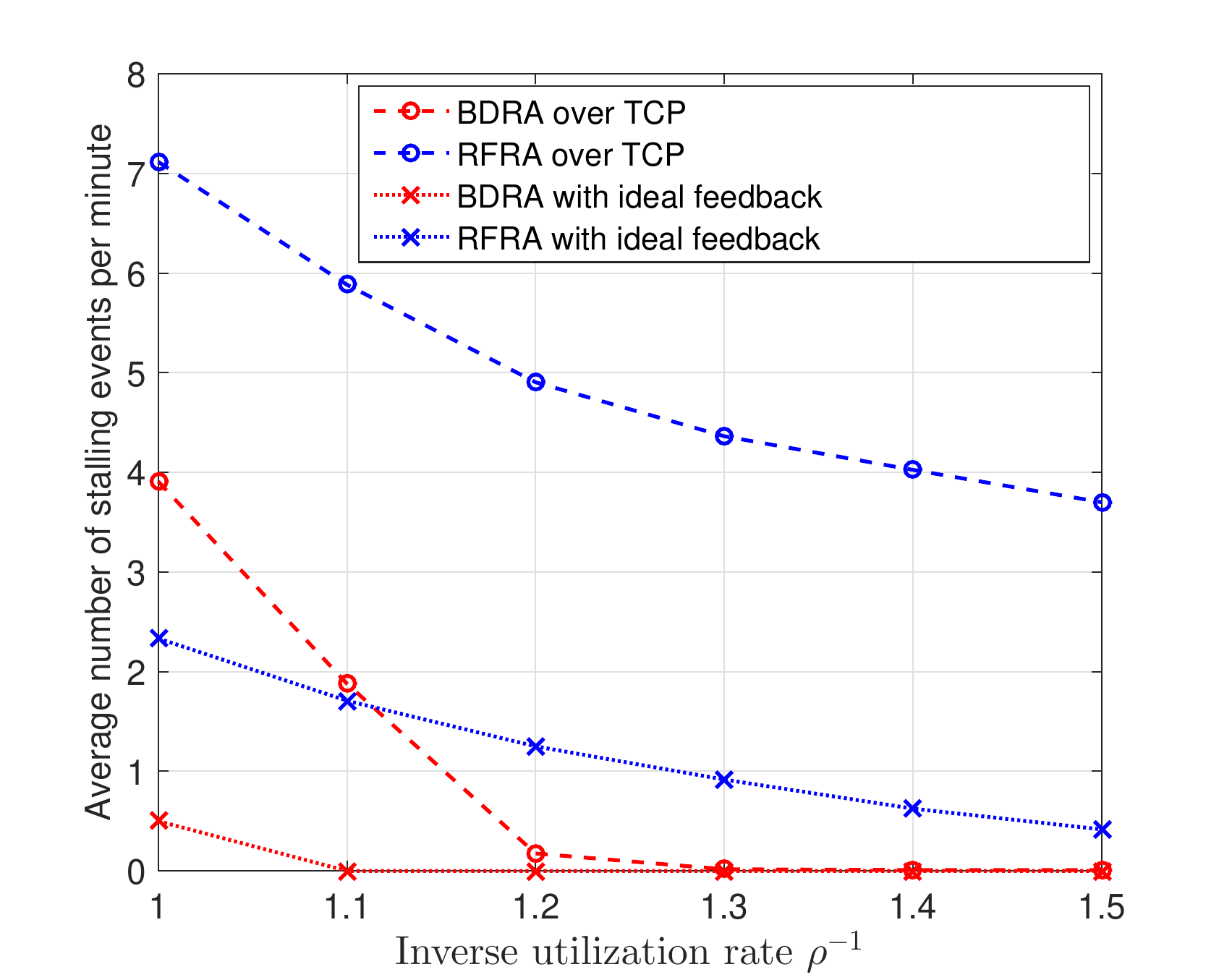}
        \caption{Video duration is $8\times16/15$ minutes.}
    \end{subfigure}
    \begin{subfigure}[b]{0.45\textwidth}
        \includegraphics[scale=0.45]{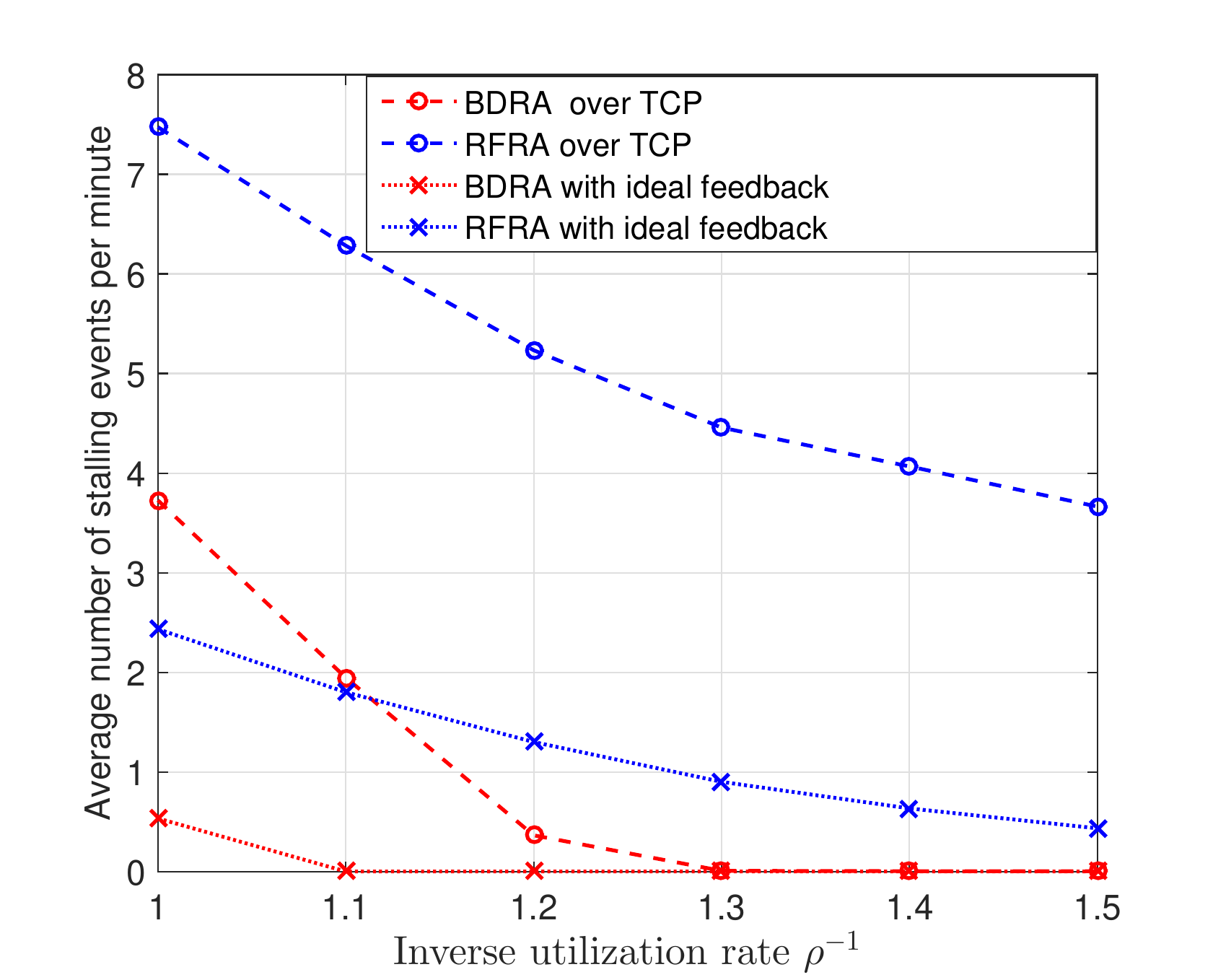}
        \caption{Video duration is $10\times16/15$ minutes.}
        \end{subfigure}    
    \caption{Average number of stalling events per minute versus inverse utilization rate $\rho^{-1}$.}\label{sec1}
\end{figure*}
\subsection{Average Number of Stalling Events per Minute}
In this section, we investigate the average number of stalling events per minute with respect to the network utilization rate and video duration. The average number of stalling events per minute is defined as the ratio of the total number of segment stalling events of all users and the total number of users multiplied by the video duration.  In our simulations, we assume that the clients have infinite size buffers used for storing incoming video packets.  Whenever a stalling event occurs, the client freezes the display of the video through  a certain prescribed time duration.  
 \begin{figure*}
    \centering
         \begin{subfigure}[b]{0.45\textwidth}
        \includegraphics[scale=0.4]{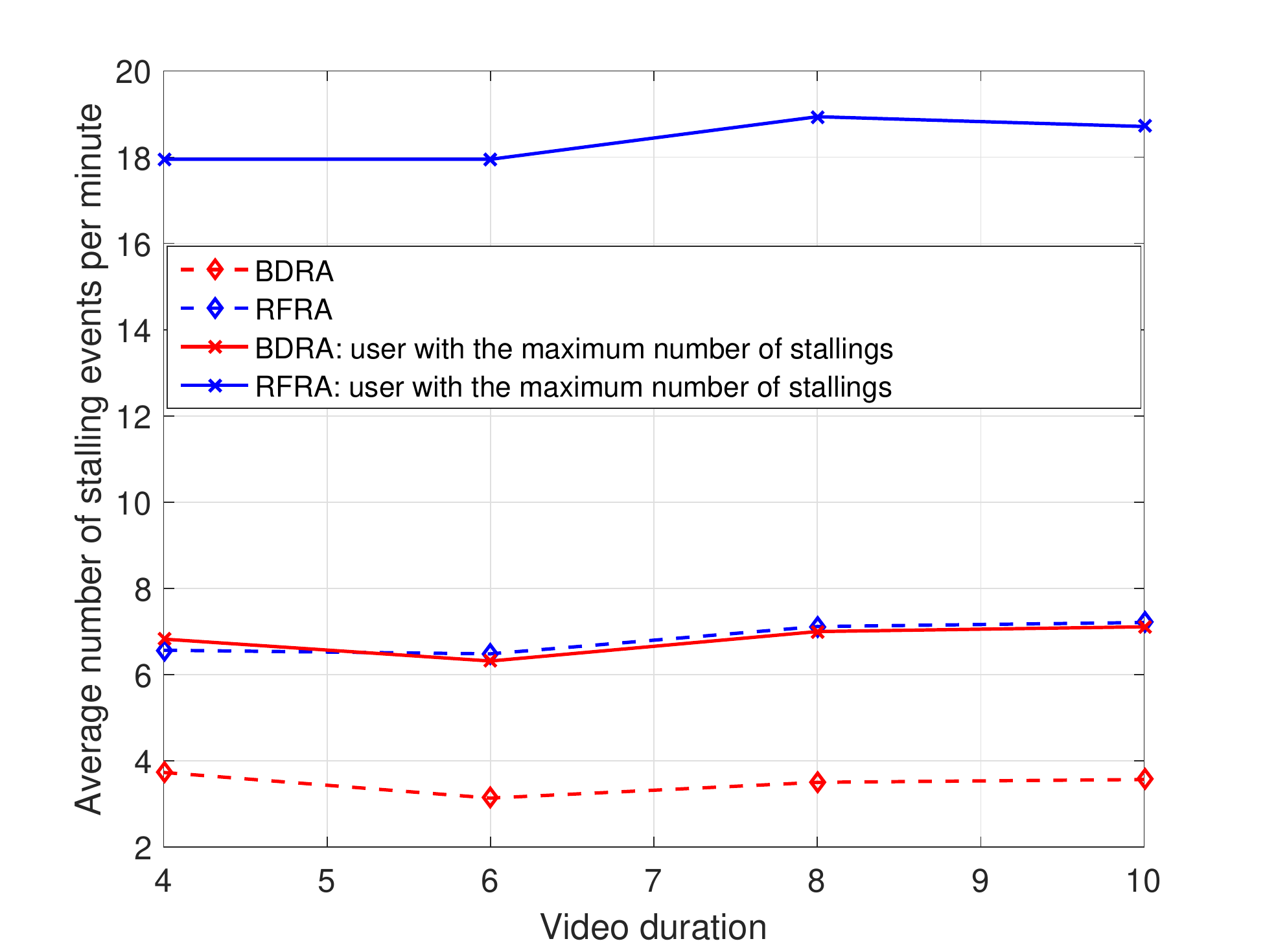}
        \caption{Rebuffer duration is 2 seconds, and $\rho^{-1}=1$.}
    \end{subfigure}
    \begin{subfigure}[b]{0.45\textwidth}
        \includegraphics[scale=0.4]{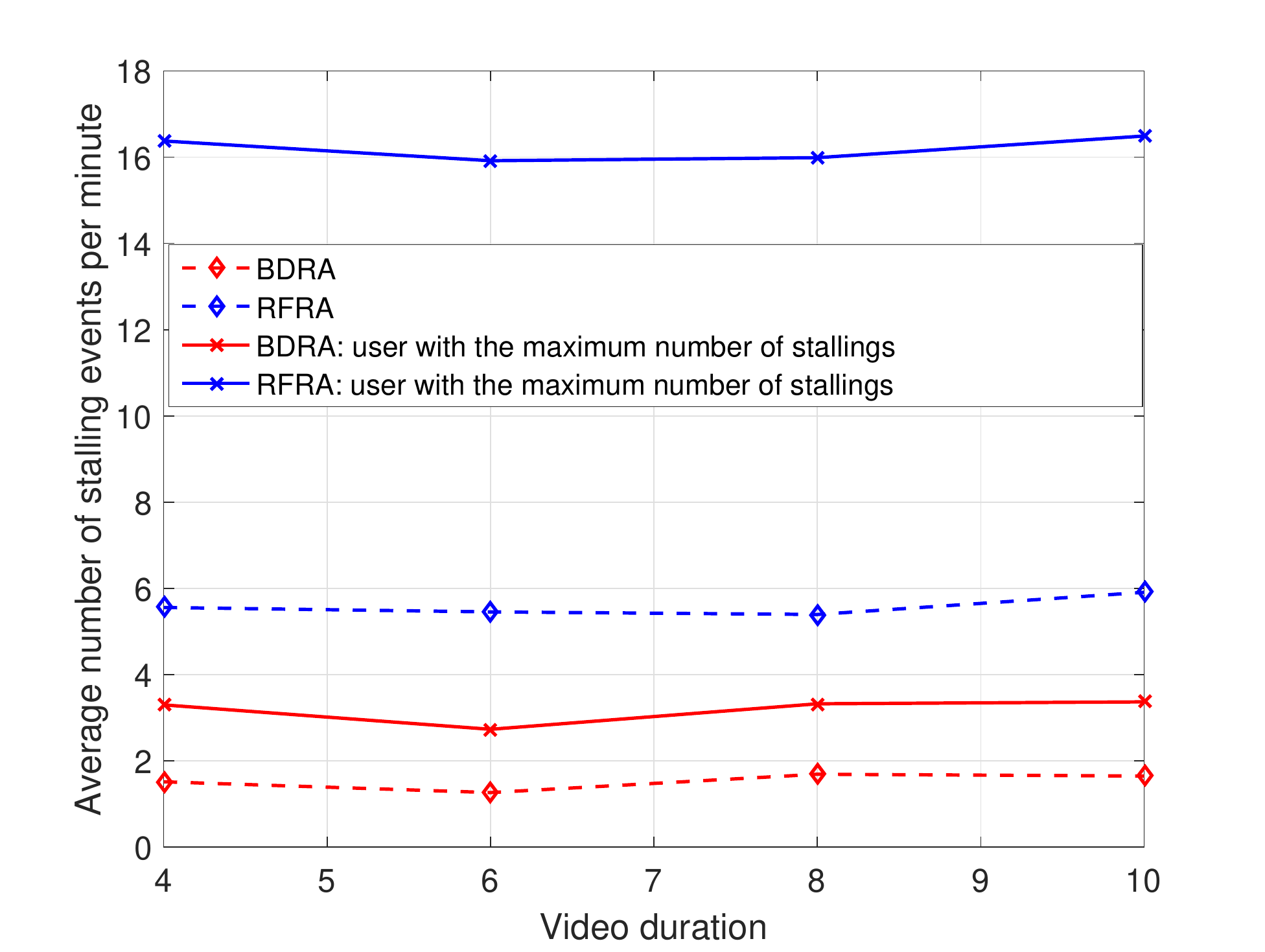}
        \caption{Rebuffer duration is 2 seconds and $\rho^{-1}=1.1$.}
        \end{subfigure}
         \begin{subfigure}[b]{0.45\textwidth}
        \includegraphics[scale=0.4]{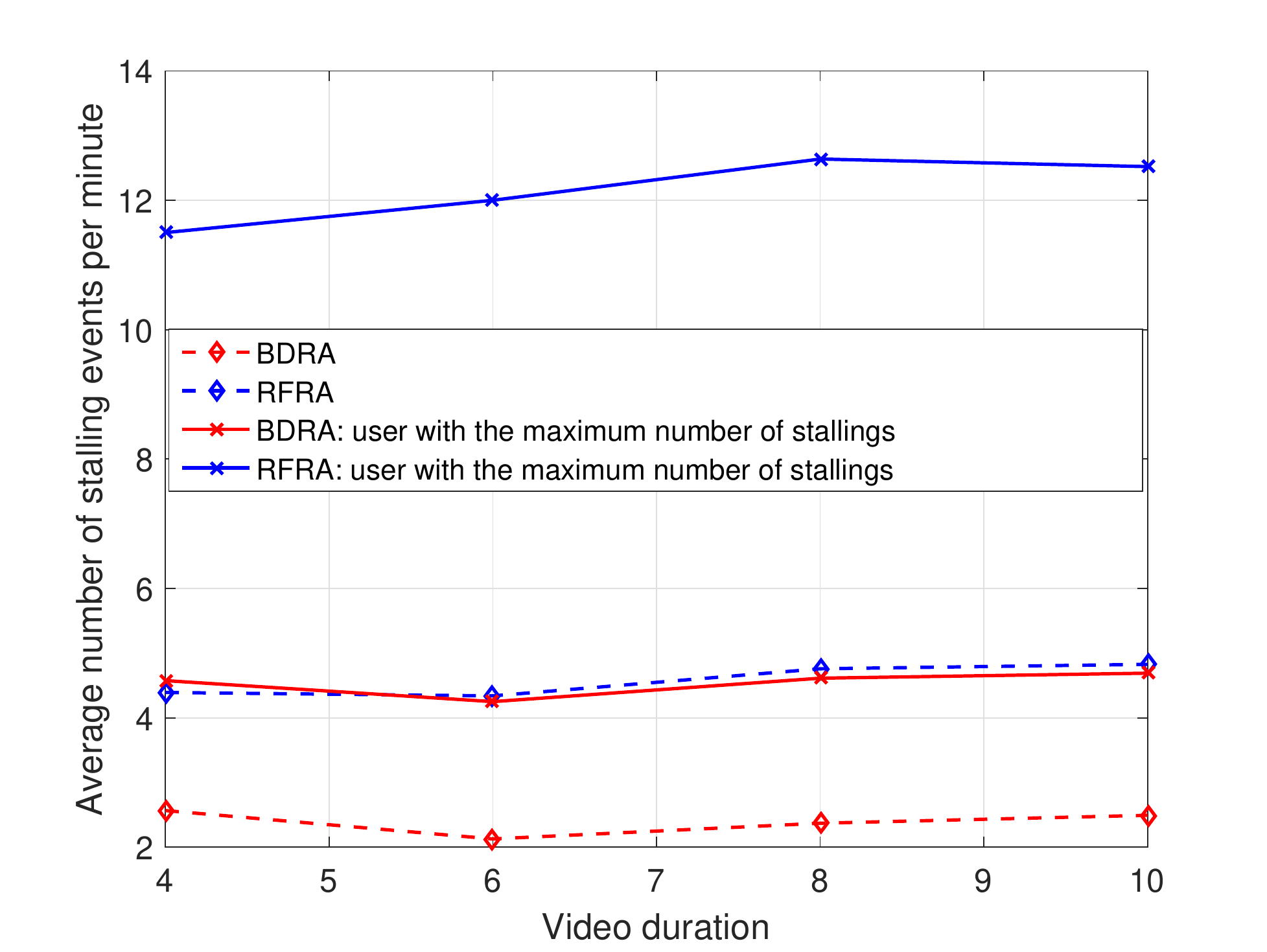}
        \caption{Rebuffer duration is 3 seconds and $\rho^{-1}=1$.}
    \end{subfigure}
    \begin{subfigure}[b]{0.45\textwidth}
        \includegraphics[scale=0.4]{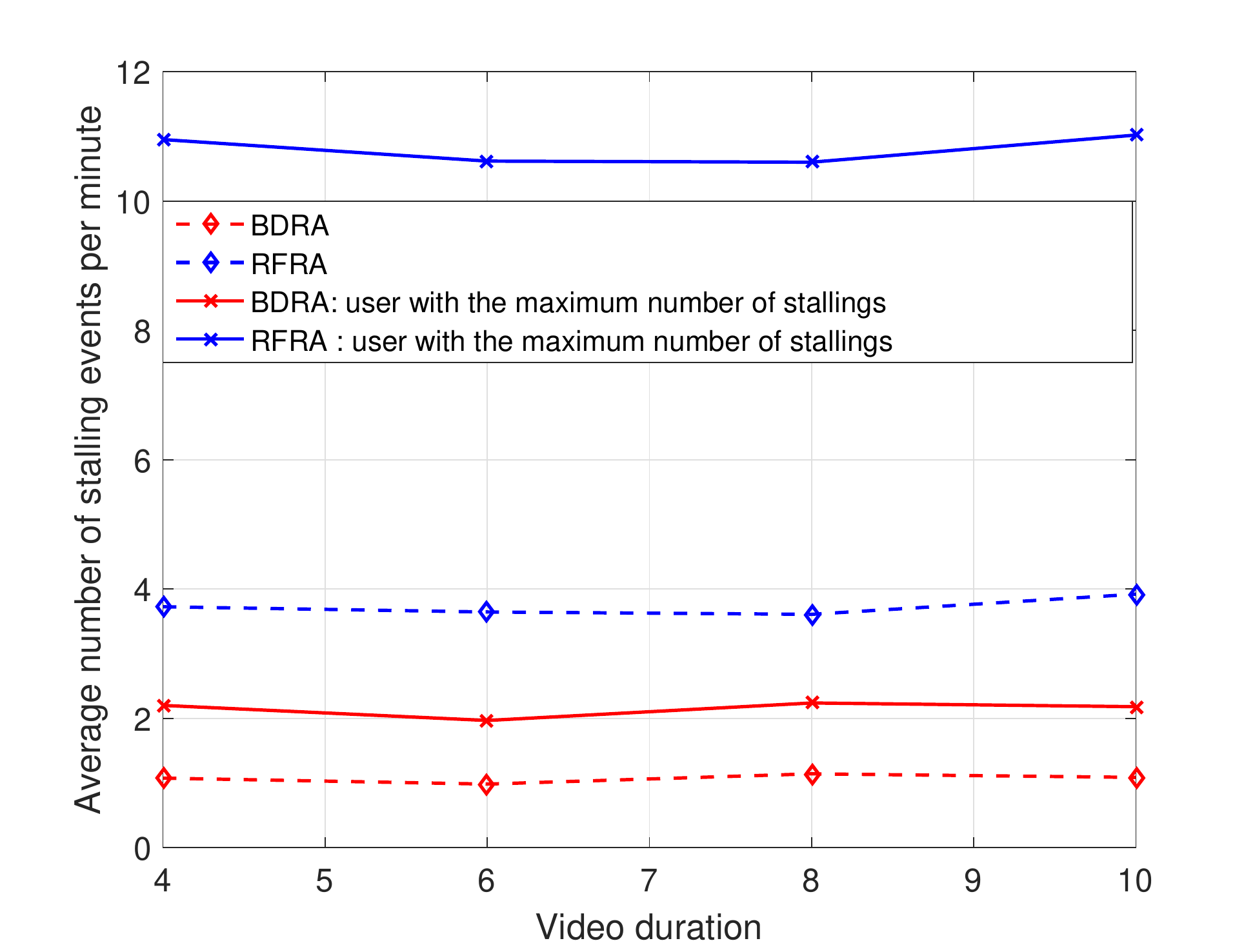}
        \caption{Rebuffer duration is 3 seconds and $\rho^{-1}=1.1$.}
    \end{subfigure}
         \begin{subfigure}[b]{0.45\textwidth}
        \includegraphics[scale=0.4]{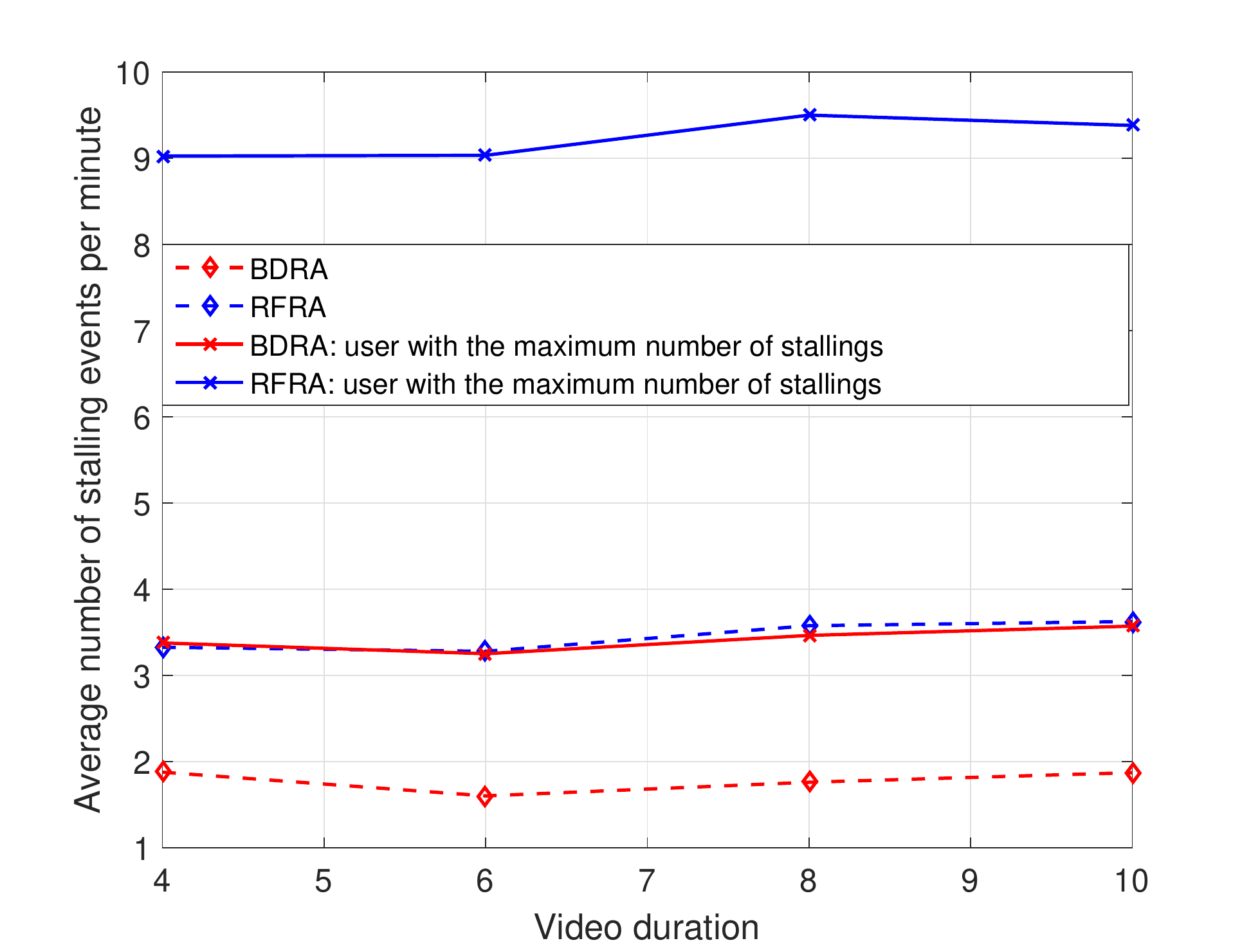}
        \caption{Rebuffer duration is 4 seconds and $\rho^{-1}=1$.}
    \end{subfigure}
    \begin{subfigure}[b]{0.45\textwidth}
        \includegraphics[scale=0.4]{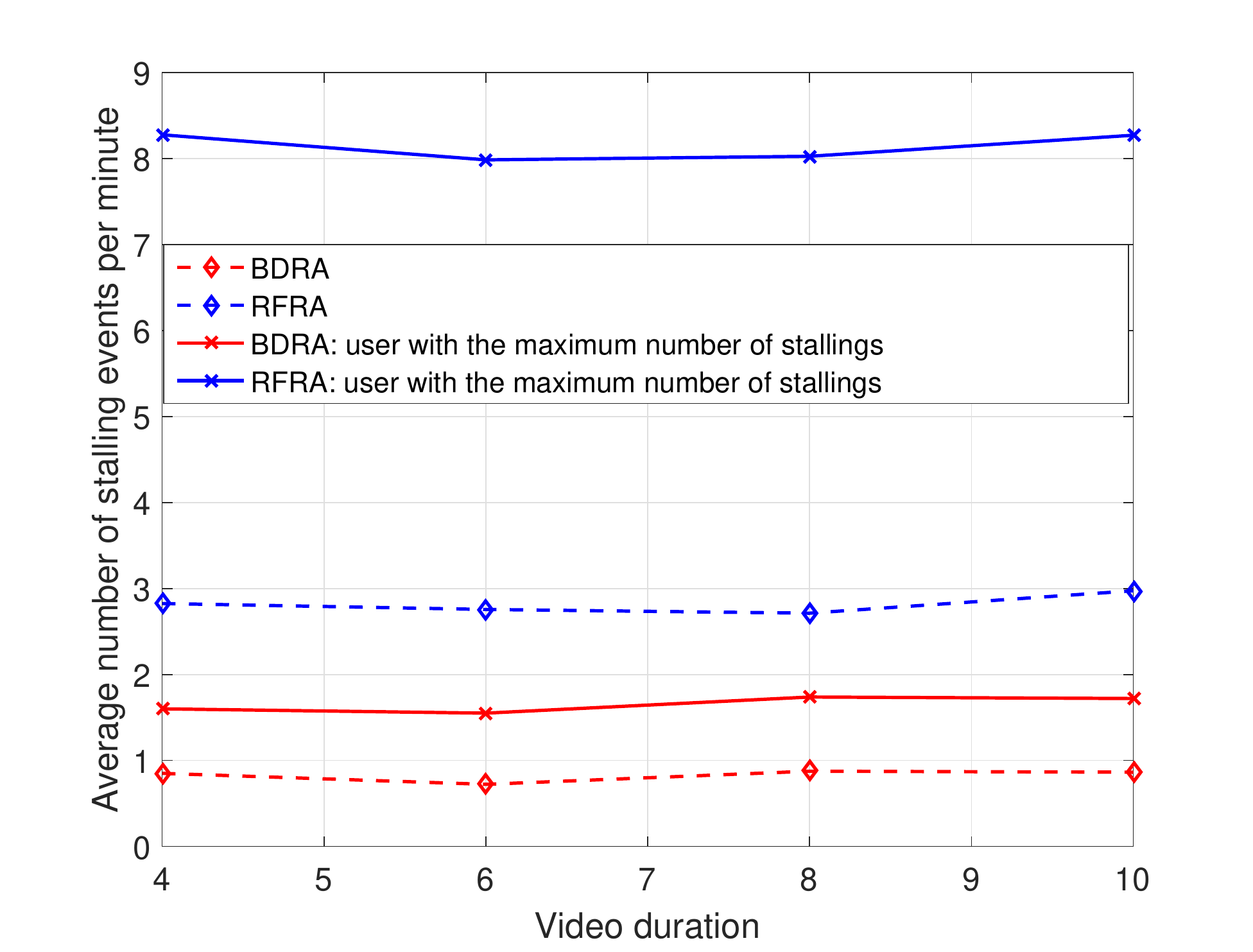}
        \caption{Rebuffer duration is 4 seconds and $\rho^{-1}=1.1$.}
    \end{subfigure}
    \caption{Average number of stalling events per minute with respect to video duration (scaled by 15/16) and rebuffer duration.}\label{sc2}
\end{figure*}
The performance of the BDRA scheme is compared with that of the RFRA  scheme which is in-cognizant of the temporal properties of the video file.  In particular, the RFRA scheme allocates communication resources to users equally in a time division manner, i.e., channel resources are allocated to users sequentially until certain number of packets is transmitted. 


We first set the \emph{initial buffer} duration to 0 second, which means that the client starts displaying the video immediately upon the arrival of the first video packet.  The \emph{rebuffer} duration is set to be 2 seconds, which means that the client freezes the display of the video during 2 seconds upon a stalling event. There are three clients, where each requesting the video files for Tokyo Olympics $QP=10$, Star Wars IV $QP=10$, NBC News $QP=10$ respectively. We vary the length of the videos between $4\times16/15$ and $10\times16/15$ minutes (i.e., 24 to 60 segment), and for each duration we vary the value of inverse utilization rate $\rho^{-1}$ from 1 to 1.5 in steps of 0.1. Each experiment scenario is repeated for 10 times with different random seeds for the random channel loss model. The results presented are the averages of these experiments.

As illustrated in Fig. \ref{sec1}, the average number of stalling events per minute decreases and ultimately approaches to zero as the channel data rate becomes much higher than the total requested video rate (i.e., as the inverse utilization rate increases).  Note that the average number of stalling events per minute with the BDRA scheme is at least $1.75$ times lower than that of the RFRA scheme, when TCP transport layer protocol is used.  Also, as the inverse utilization rate increases, the average number of  stalling events per minute decreases much rapidly for the BDRA scheme. 
 In fact, the BDRA scheme can provide service with no stallings if the inverse utilization rate is more than $1.3$ when implemented with TCP and $1.1$ when implemented with an ideal transport layer. The RFRA scheme cannot provide service with no stallings when the inverse utilization rate is less than $1.5$ for both implementations (TCP and ideal transport layer). The RFRA algorithm is much more adversely affected by the TCP implementation than the BDRA scheme, with its average stalling events per minute remain in the range of four stallings per minute even when inverse utilization rate is more than $1.5$. We also observe that the video length has almost no effect on the outcome of the experiments for both schemes.

The transport layer has significant impact on the performance. In the simulations, both algorithms are implemented at the application layer, and they wait until certain number of packets are send to switch to another user.  Once there is a packet loss, the TCP time-out mechanism is provoked if an ACK is not received after \emph{Retransmission Timeout (RTO)} duration. As per the specifications given in RFC6298 \cite{rfc6298}, the minimum RTO duration is $1$ second, even though this duration can be optimized to improve efficiency \cite{optrto}.  Note that during an RTO duration, no new packets are sent and the link becomes under-utilized.  This affects not only the ongoing transmission, but also the subsequent GoP transmissions to other clients by limiting the amount of time that can be used to deliver video packets before their deadlines.

In the next set of simulations, we investigate the effect of rebuffer duration and video length. We implement the obtained BDRA scheme  together with TCP layer only.  We set the initial buffer duration to $4$ seconds, and the rebuffer durations are taken $2$, $3$, and $4$ seconds. The duration of the video is $\left\{ 4, 6, 8, 10\right\}\times16/15$ minutes.  We performed the simulations for inverse utilization rates $\rho^{-1}$ of 1 and 1.1. And each experiment scenario is repeated for 10 times with different random seeds for the random channel loss model, and we then take average of them. Fig. \ref{sc2} indicates that the average number of stalling events per minute stays approximately the same with increasing video length for a given rebuffer duration and $\rho^{-1}$.  We also observe that the rebuffer duration is another important factor for decreasing the number of stalling events with its impact more prominent for larger $\rho^{-1}$ values. Also note that the improvement in video stalling events is more significant when the rebuffering duration is increased from $2$ seconds to $3$ seconds, but this improvement gets smaller for higher rebuffer durations.  

In Fig. \ref{sc2}, we depict the average number of stallings events per minute averaged over all clients, as well as for the client which has the highest the number of stallings among the three in the network.  Although it is not identified as one of our main initial objective, we observe that the BDRA scheme performs much more fairly than the RFRA scheme in this aspect, too. The performance of the worst performing client, who requests the video stream with the highest source rate, is much closer to the average performance in the network as compared to that with the RFRA scheme.
  \begin{table}
  \caption{Comparison of WRFRA scheme with BDRA scheme}\label{WRR}
\centering
  \begin{tabular}{|M{1cm}|M{3cm}|M{3.5cm}|}
    \hline
$\rho^{-1}$ & \textbf{Resource Allocation Scheme} & \textbf{Average Number of Stallings per Minute per User} \\ \hline
    \multirow{3}{*}{1.1} & BDRA(ideal)  & 0.531 \\
		& BDRA(Super-GoP)  & 1.781 \\
     & WRFRA  & 6.062 \\ \hline 
    \multirow{3}{*}{1.2} & BDRA(ideal) &  0.0218\\
		& BDRA(Super-GoP) &  0.025\\
     & WRFRA & 4.312\\ \hline 
		\multirow{3}{*}{1.3} & BDRA(ideal) &  0\\
		& BDRA(Super-GoP) &  0\\
     & WRFRA  & 1.937\\ \hline 
  \end{tabular}
\end{table}

 \begin{table*}
   \caption{Simulation results for quality adaptation}\label{adaptiveresults}
  \centering
  \begin{tabular}{|M{1cm}|M{4cm}|M{2cm}|M{3cm}|M{3cm}|}
    \hline
    $\rho^{-1}$ & \textbf{Resource Allocation Scheme} &  \textbf{Number of Stallings per User}  &  \textbf{Average Quality}    &  \textbf{Average Quality of the Worst User} \\ \hline
    \multirow{4}{*}{1} & BDRA(ideal) & 11 & 5.858  & 5.675\\
		& BDRA(Super-GoP) & 14 & 5.785 & 5.666\\
		 & RFRA allocation & 33.666 & 5.758  & 5.358\\ 
     & WRFRA & 34.666 & 5.733  & 5.675\\ 
		& DWRFRA & 20.33 & 5.83  & 5.808\\ \hline 
    \multirow{4}{*}{1.1} & BDRA(ideal) & 5.666 & 5.941  & 5.85\\
		& BDRA(Super-GoP) & 7.333 & 5.922  & 5.883\\
		& RFRA & 28 & 5.791  & 5.408\\
     & WRFRA& 26.666 & 5.819  & 5.775\\ 
		& DWRFRA & 10 & 5.9  & 5.875\\ \hline
  \end{tabular}
\end{table*}
\subsection{Comparison with Weighted Rate-Fair Resource Allocation}
 In our simulations above, we only considered the RFRA scheme.  Recall that the clients are served with equal average rates in the RFRA scheme. However, when the clients request video files with different bit-rates, their rate requirements will not be equal. Hence, the RFRA scheme does not guarantee a fairness among users in terms of the QoE. In that case, the resource allocation to users can be performed in proportion to the bit-rate requirements of the requested video files.

 We call this type of resource allocation scheme {\em weighted rate-fair resource allocation} (WRFRA) . While we still expect the BDRA scheme to perform better than the WRFRA scheme since the BDRA scheme also takes bit-rate fluctuations into account, the WRFRA scheme is expected to perform better than the RFRA scheme.  To test this hypothesis, we conducted the second set of simulations with WRFRA scheme for video files with duration $10 \times 16/15$ minutes (i.e., 120 segments) and initial buffer and rebuffer durations are set to $3$ and $2$ seconds respectively. We remark that average bit-rates of the requested video files are approximately proportional to the weights of $2$, $1$ and $4$. Hence, the WRFRA scheme allocates the channel resources to users according to these weights.\\
\indent In the simulations, we also implement the BDRA scheme with super-GOP approach, where the clients send HTTP-GET request for only segments (super-GOPs) instead of each GoP, in addition to ideal BDRA scheme. In the super-GOP implementation, the access point update the deadlines of the packets when there is a request for the next segment. Since the access point is not able to identify the GoPs in the requested segment in this case, we use a deadline for the segment instead of each GoP. The deadline of each segment is set to the deadline of the first GoP of the corresponding segment\\ 
\indent  Table \ref{WRR} shows the statistics for stalling frequency both with the BDRA scheme and WRFRA scheme. It can be observed that such a weight assignment improves the performance of the RFRA scheme, but the BDRA scheme  still performs much better than the WRFRA scheme, especially under heavily loaded network conditions. Indeed, our BDRA scheme can be considered as a WRFRA scheme with weights updated dynamically throughout the video streaming session in an optimal way. One can also observe that when $\rho^{-1}>1.1$,  the ideal BDRA scheme and the super-GOP based BDRA scheme perform very close, however when  $\rho^{-1}=1.1$  the difference between the performances of two different implementation of the BDRA scheme is more visible. Hence, this observation simply implies that the deadline information is more critical when the network is highly loaded.

\subsection{BDRA Scheme with Video Quality Adaptation}
\label{sec: adaptive}

 In order to assess the performance of the BDRA scheme with adaptive bit-rate streaming, we also implemented a simple bit-rate selection mechanism along with the BDRA scheme. Our results are summarized in Table \ref{adaptiveresults}. The basic simulation set-up to obtain the results in this table is similar to those above: $3$ users request the video files Tokyo Olympics $QP = 10$, Start Wars IV $QP = 10$ and NBC News $QP = 10$ with durations of $10 \times 16 / 15$ minutes (i.e., 120 segments) and initial buffer and rebuffer durations are set to $3$ and $2$ seconds, respectively. For the video quality adaptation, we consider $L=6$ different quality levels, i.e., $\left\{1,2,3,4,5,6\right\}$, which corresponds to different resolutions, i.e., $240$p, $360$p, $480$p, $720$p, $1080$p and $1440$p. The bit-rate of the corresponding resolutions/quality levels are proportional to weights $Q_{1}=0.05$, $Q_{2}=0.08$, $Q_{3}=0.13$, $Q_{4}=0.26$, $Q_{5}=0.47$ and $Q_{6}=1$. For the video quality adaptation, we assume that the  quality level $6$ indicates the same bit-rate with the original video file and quality level $l<6$ indicates the bit-rate level which is equal to the bit-rate of the original video file weighted by $Q_{l}$.\\ 
 \indent The switching mechanism between quality levels is as follows. If a user experiences an interruption in its video streaming service while on quality level $l>1$, then the next segment is requested at the quality level $l-1$. On the other hand, if a user does not experience any stalling event in the current
segment while on quality level $l<6$ and the lead time between the last GoP delivery time and the deadline is above a threshold value (i.e., we take this threshold value as $3$ GoP durations in
our simulations), then the user requests the next segment with a higher quality level of $l+1$. We note that, although the quality level is decreased when there is a stalling event, a threshold based policy, like the one used for the quality level increment, can be employed to pro-actively reduce the quality level of the streaming video in order to prevent a possible jitter event in the future. We also want to remark that we assume the same threshold value for each quality level increment step.  However, through more advanced ways of selecting threshold values (e.g., assigning higher threshold values to higher quality levels or dynamically assigning threshold values according to $\rho^{-1}$ and the number of users), the algorithm performance can even be improved further. This direction is not within the scope of this paper and will be considered as a future work. For the numerical results, we also consider a metric representing the average quality level of video streaming sessions, which is defined according to  
\begin{equation}
 \sum_{i=1}^{N} \sum_{l=0}^{L-1}l\frac{S_{i,l}}{N S_i},
\end{equation}
where, for $i \in \{1, \ldots, N\}$,  $S_{i,l}$ and $S_{i}$ are the number of segments received by user $i$ at quality level $l\in\left\{1,\ldots,6\right\}$ and the total number of segments received by user $i$, respectively.  

We consider the performance results for each value of $\rho^{-1}\in\left\{1, 1.1\right\}$. For each $\rho^{-1}$, we obtain three different performance metrics: (i) the average number of stalling events per user, (ii) the average quality level and (iii) the average quality level of the worst case user. We compare $5$ different blind resource allocation schemes. These are the BDRA scheme with GoP requests, the BDRA scheme with segment requests (i.e., super-GoP requests), RFRA, WRFRA  and dynamic weighted rate-fair resource allocation (DWRFRA). The DWRFRA scheme is an extension of the WRFRA scheme where the weights dynamically change according to GoP sizes. The fundamental difference between the DWRFRA scheme and the BDRA scheme is that the BDRA scheme further utilizes the deadlines of the GoPs and prioritizes users with the small GoP sizes to minimize the number of stalling events. Performance comparison of these $5$ resource allocation algorithms is illustrated in Table \ref{adaptiveresults}.

As expected, we observe that DWRFRA outperforms the WRFRA algorithm in all performance metrics since DWRFRA changes the weights dynamically.  We also observe that although the RFRA scheme achieves higher average quality level compared to the WRFRA when $\rho^{-1}=1$, the average quality of the worst case user is higher in the case of WRFRA. The root cause for this observation is that the RFRA scheme favors the users requesting low bit-rate files, which leads to a deterioration in the performance of the users requesting high bit-rate files. Hence, the observed high service quality of the RFRA scheme stems from the selection of higher quality levels for low rate videos.

In terms of the fairness, DWRFRA, WRFRA and BDRA schemes obtain a balanced average quality selection over all users (i.e., average quality level is close to the average quality level of the worst case user) without biasing any user predominantly unlike RFRA. From this perspective, the BDRA scheme can be thought to be more fair than the plain RFRA scheme. An interesting observation is that the highest average quality level of the worst case user is achieved by the DWRFRA scheme. We remark that the BDRA scheme prioritizes the GoPs with smaller sizes.  Hence, the clients streaming higher bit-rate video files are more prone to stalling events.  Although this design leads to unfairness between the clients in terms of the received video quality level, it still implies certain level of fairness in terms of the bandwidth usage. In addition, one can observe from Table \ref{adaptiveresults} that less number of stalling events is observed with the BDRA scheme when compared to the DWRFRA scheme by virtue of its prioritization. Hence, when the clients are streaming video files with different bit-rates, there is a trade-off  between the number of system-wide stalling events and the average quality level of the worst case user. 

Finally, we also observe that with increasing $\rho^{-1}$,  the system performance improves faster with the BDRA scheme when compared to the RFRA and WRFRA schemes according to all three quality measures (i.e., number of stallings per user, average quality, and the average quality of the worst case-user). These benefits of the BDRA scheme are due to its two main properties. First, the BDRA scheme strives to improve the system-wide stalling frequency performance through its deadline based structure. Hence, it does not favor any one of the users requesting video streaming service since such a biasing will degrade the collective system performance. Second, our implementation of the BDRA scheme, as explained above, prioritizes the users with smaller GoP sizes. This implementation helps the BDRA scheme to prevent high bit-rate files from overwhelming the network, as observed through more balanced quality selection and service interruptions over all users.                 

\section{Related Work}
\label{sec:related}
\subsection{Adaptive Video and DASH}
There is a plethora of work on the adaptation of quality of video with respect to the network conditions.  In particular, many prior studies investigate the use of scalable video coding (SVC) for this purpose \cite{svc1,svc2,svc3,svc4,svc5,svc6}.  In almost all of these works,  a rate-distortion metric is first constructed by considering the specific structure of the video. Then, a utility function, which is defined with respect to this metric, is optimized by developing different scheduling, rate allocation, and admission control policies.  Although there is still an ongoing interest on the use of SVC along side DASH \cite{svcdash}, in most commercial applications this alternative is forgone, mainly because SVC requires a control mechanism at the server side that introduces an added complexity contradicting with the initial premise of the DASH structure.

In \cite{dash1,dash2,dash4}, the process of video streaming over HTTP is explained in detail and the specifications of the DASH protocol are introduced. The rate-adaptation mechanisms of commercial players are investigated in \cite{exp1}. After the standardization of DASH protocol, a great deal of attention is devoted to the client side control algorithms for bit-rate (quality) selection in order to maximize the video quality, and to minimize the  fluctuations in quality and the number of stalling events \cite{csca1,csca2,csca3,csca4,csca5,csca6}. 
The main purpose of all these client-centric algorithms is to adjust the video source rate of users according to available resource in order to prevent congestion and corresponding stalling events. However, in our work, we concentrate on the time period between the two bit-rate selection instants and show that it is possible to reduce the number of stalling events further via implementing an additional  low-complexity scheduling algorithm at the server side.     

\subsection{Client-Side Approaches}

It is reported that client-side approaches give rise to other problems due to over/under-estimation of the actual available bandwidth. In \cite{compete1}  and \cite{compete2}, the authors have identified three main performance issues: (i) player instability, (ii) unfairness between the players and (iii) the under-utilization of the available bandwidth. It is pointed out that the main cause of these performance issues is the successive activity and inactivity periods, which leads to miscalculation of the available bandwidth.  In line of this work, other studies have investigated the efficient and fair utilization of the available bandwidth \cite{cscax,compete3,compete4,compete5}.

In \cite{cscax}, two control mechanisms are implemented at the client side; one for controlling the playout buffer, and one for selecting the appropriate video quality level that matches the best-effort bandwidth. Furthermore, two actuators are implemented at the server; one changing the video quality, and the other throttling the video streaming rate. The fairness issues are addressed in \cite{compete3} and \cite{compete4}. In \cite{compete3}, the authors aim to adjust the rate of each segment to eliminate the off-periods (time period during which the client stops requesting a video segment). It was conjectured that the elimination of the off-periods improves the accuracy of  TCP-based bandwidth estimation procedures. 
In \cite{compete4},
a randomized segment scheduler is used to prevent the problem of overestimation/underestimation of the available bandwidth due to a biased view of the network state.  Li {\em et al.} \cite{compete5} proposed {\em probe and adapt (PANDA)} method to improve bandwidth utilization. PANDA probes the available bandwidth via an {\em additive increase and multiplicative decrease} (AIMD) based method over a segment cycle in order to prevent misleading bandwidth estimation. Proposed algorithms in \cite{compete3,compete4,compete5} aim to increase accuracy of the resource estimation process, and unlike our method introduced in the paper, these  approaches are client centric. 

\subsection{Server-Side Approaches}
 In \cite{proxy1}, a proxy based traffic and resource management framework for LTE networks is introduced. In this framework, a QoE optimizer is used to obtain the optimal transmission rates of the clients which maximize the aggregate video utility. These rates are utilized by the LTE scheduler as the target transmission rates. Further, the proxy may overwrite the clients' segment requests according to the feedback signals received from the QoE optimizer and the buffer levels of the clients in order to maximize the aggregate video utility. A similar framework is considered in \cite{proxy2},  but with the performance objective being a fair QoE maximization rather than being the aggregate video utility maximization as in \cite{proxy1}.  Similarly, network assisted video quality assignment and bandwidth allocation schemes for improving QoE in HTTP video streaming systems are also studied in some recent papers such as \cite{channelbased, Colonnese15} and \cite{networkassist2}.  The solutions in these papers require various side information to run properly such as network throughput rates, channel quality indicators, clients' instantaneous buffer states and clients' buffer occupancy  trends.  On the other hand, the BDRA scheme we derive in this paper operates without requiring any such information at the server-side and minimizes the system-wide stalling probability under such no-feedback conditions.
 
 
\begin{table*}[t]
\caption{Summary of related work and comparison with the derived BDRA algorithm.}
\label{Table: Literature Review}
\centering
\begin{tabular}{M{1.95cm}|M{1.95cm}|M{2.8cm}|M{1.95cm}|M{1.95cm}|M{1.95cm}|M{1.95cm}|}
\cline{2-7}
                       & \multicolumn{2}{c|}{\textbf{Algorithm Objective}} & \multicolumn{4}{c|}{\textbf{Required Side Information}} \\ \cline{1-7}
\multicolumn{1}{|c|}{Algorithm} & Resource Allocation       &  Bit-Rate Guidance         & Client Buffer Status   &  Channel State Information  &  Rate-Based Utility Function  &   Blind (only HTTP-GET Requests)    \\ \hline
\multicolumn{1}{|c|}{\cite{proxy1,proxy2,networkassist2,future}} & $\times$ &  $\times$ &  $\times$ &  $\times$ & $\times$ &   \\ \hline 
\multicolumn{1}{|c|}{\cite{networkassist},\cite{sdn1}} &  $\times$ &  $\times$ &   &  & $\times$ &        \\ \hline
\multicolumn{1}{|c|}{\cite{main2}} & $\times$ &    &  $\times$ &  $\times$ &  &          \\ \hline
\multicolumn{1}{|c|}{\cite{bufferaware}} &  $\times$ &  $\times$  & $\times$ &   &  &          \\ \hline
\multicolumn{1}{|c|}{\cite{sdn2}} &   $\times$ & $\times$   & $\times$ &   & $\times$ &          \\ \hline
\multicolumn{1}{|c|}{\cite{channelbased}} &   $\times$ & $\times$   &  & $\times$  &  &          \\ \hline
\multicolumn{1}{|c|}{BDRA - Our Solution} &   $\times$ &    &  &   &  & $\times$ \\ \hline
\end{tabular}
\end{table*}

Last but not least, there are also other recent papers that customize the software-defined-networking (SDN) architecture for the HTTP adaptive streaming scenarios in order to have a central controller helping users to select the optimal video quality level and helping the access point to share available resources among the users intelligently \cite{sdn1, sdn2, bufferaware, networkassist}. In \cite{bufferaware}, the authors propose a guidance mechanism for distributing HTTP-GET requests over time in order to improve QoE in HTTP video streaming systems. Their solution depends on the accurate estimation of data rate fluctuations over time, which can be an onerous task for when channel conditions change rapidly as in the wireless communication environments. Different from \cite{bufferaware}, our BDRA scheme is not restricted by the estimation process of randomly varying network conditions over time thanks to its blind operation. The papers \cite{sdn1,sdn2} aim to improve QoE of users via jointly optimizing the resource allocation and video quality levels on a segment scale. Different from them, we are mainly interested in a GoP level optimization that spans the time interval between two segment requests in this paper. To this end,  we obtain a  deadline-based rate allocation policy that can track the bit-rate fluctuations of the video files without requiring an additional feedback mechanism on top of the usual HTTP-GET requests available in a DASH based video streaming system.   The solution in \cite{networkassist} is based on the existence of a network hypervisor that either determines an allocation of bandwidth slices to streaming clients or guide streaming clients in their video bit-rate selection process by using its network-wide knowledge on link capacities. Our BDRA scheme complements any bit-rate guidance mechanism in the SDN setup due to its operation blind to clients' quality adaptation mechanism. The bandwidth slicing approach in \cite{networkassist} divides the available bandwidth in proportion to the bit-rates of the requested video files, which is akin to the weighted rate-fair resource allocation (WRFRA) scheme implemented in our paper. And, we show that our BDRA scheme performs significantly better than the bandwidth slicing approach in terms of the frequency of stalling events.

\subsection{Summary of the key issues}

To conclude this part, we summarize the related work that is most relevant to our study in this paper along two main dimensions of algorithm objective and required side information in Table \ref{Table: Literature Review}. As this table makes it further clear, the BDRA algorithm allocates communication resources to multiple video streaming clients by the assistance of HTTP-GET requests only, whereas the required side information to run properly is much larger for other existing HTTP-based video streaming solutions. Hence, an emergent salient feature of our solution is its being blind to instantaneous channel conditions, clients' buffer sizes and clients' video adaptation mechanisms while scheduling the users in order to minimize the stalling probability in DASH based multiuser video streaming systems. The specific fundamental differences of the derived optimum BDRA algorithm in this paper, when compared to the previous work, can be listed as (i) its operation without requiring dedicated feedback communication, (ii) its operation that is blind to the operation of clients and their experienced channel conditions, and (iii)  its operation that does not directly intervene with the quality selection process of the clients.

\section{Conclusions and Future Directions}
\label{sec:conclusion}
This work introduces a DASH compatible network assisted control mechanism to be implemented at the edge server. We have first presented the notion of optimal slot based resource allocation policy to minimize the segment stalling event probability. Then, we have analytically showed that the derived blind deadline-based rate allocation (BDRA) scheme minimizes the system-wide segment stalling probability when only average channel state information is available. 
We have demonstrated the efficacy of the algorithm with a realistic NS-3 simulation depicting its performance over an ideal transport layer with perfect feedback, as well as over a more common TCP transport layer.  The simulations also demonstrate that the BDRA scheme better utilizes the channel as compared to other rate-fair resource allocation schemes.\\

Note that the access point is oblivious to the instantaneous channel states in our model and the BDRA algorithm is proven to be optimal when instantaneous channel state information is not available.  Although it is possible to achieve a higher network performance with an opportunistic rate allocation scheme using the instantaneous channel conditions, this will induce significant overhead and complexity in the system design. However, with the BDRA scheme, only certain features of the DASH protocol is utilized, e.g., GoP structure and HTTP-GET requests. Thus, the BDRA scheme is a blind algorithm in a sense that it is executed without provisioning the client side video bit-rate adaptations. In this network model, the BDRA scheme is totally excluded from the video quality selection procedure and only aims to minimize the number stalling events over all clients for chosen video qualities. A hybrid approach, where both client and server have control over the video quality selection, may improve the network performance.

In fact, there is a recent work that investigates a hybrid control mechanism for multi-user video streaming taking advantage of the computational efficiency of cloud computing \cite{future}. Although a joint control mechanism may improve the performance of the network, it requires a complete redesign of the DASH protocol. Furthermore, a joint control mechanism induces complexity at the server-side and requires additional feedback from the clients. In particular, server should also intervene in the video quality selection of client. An intelligent blind resource allocation algorithm that provisions the client side video bit-rate adaptations with the knowledge of DASH structure and allocates slots to users in a way that not only  minimizes the segment stalling events but also forces some clients to decrease their video quality in favor of the overall network performance can be considered as an interesting extension of our work. 




%
%
\balance
\bibliographystyle{IEEEtran}
\bibliography{IEEEabrv,MCRef}



\end{document}